\documentclass[letterpaper,12pt]{article}   
\usepackage{osajnl2} 

\usepackage{cite}
\usepackage{amsmath,epsfig,multirow,tikz}
\usepackage{epstopdf,color,booktabs,comment}
\usepackage{tikz}
\usepackage{pgfplots}
\usetikzlibrary{spy,calc}
\usepackage{hyperref}

\usepackage[caption=false,font=footnotesize]{subfig}
\usepackage{algorithm}
\usepackage{algorithmic}
\usepackage{amsmath,amsfonts,amsthm,amssymb}
\usepackage{cleveref}

\usepackage{steinmetz}

\usepackage{animate}

\DeclareMathOperator*{\argmax}{arg\,max}

\newcommand{\aopt}[0]{\underline{a}_{\sf opt}}

\newcommand{\tnorm}[0]{\tau_{\sf normal}}
\newcommand{\topt}[0]{\tau_{\sf opt}}

\newcommand{\ie}{\textit{i.e. }}
\newcommand{\etal}{\textit{et al. }}

\newcommand{\aleftp}[0]{\underline{a}^{+}_{1}}
\newcommand{\aleftm}[0]{\underline{a}^{-}_{1}}
\newcommand{\arightp}[0]{\underline{a}^{+}_{2}}
\newcommand{\arightm}[0]{\underline{a}^{-}_{2}}

\newcommand{\Nlay}[0]{N_{\sf lay}}

\newcommand{\Ft}[0]{\widetilde{F}}
\newcommand{\Diag}[0]{\mbox{diag}}
\newcommand{\Imag}[0]{\mbox{Im}}

\begin{document}

\title{The transmission coefficient distribution of highly scattering sparse random media }


\author{Curtis Jin, Raj Rao Nadakuditi, Eric Michielssen}
\address{(jsirius@umich.edu, rajnrao@umich.edu, emichiel@umich.edu}

\begin{abstract}
We consider the distribution of the transmission coefficients, \textit{i.e.} the singular values of the modal transmission matrix, for 2D random media with periodic boundary conditions composed of a large number of point-like nonabsorbing scatterers. The scatterers are placed at random locations in the medium and have random refractive indices that are drawn from an arbitrary, known distribution. We construct a randomized model for the scattering matrix that retains scatterer dependent properties essential to reproduce the transmission coefficient distribution and analytically characterize the distribution of this matrix as a function of the refractive index distribution, the number of modes, and the number of scatterers. We show that the derived distribution agrees remarkably well with results obtained using a numerically rigorous spectrally accurate simulation. Analysis of the derived distribution provides the strongest principled justification yet of why we should expect perfect transmission in such random media regardless of the refractive index distribution of the constituent scatterers. The analysis suggests a sparsity condition under which random media will exhibit a perfect transmission-supporting universal transmission coefficient distribution in the deep medium limit.
\end{abstract}


\maketitle 

\section{Introduction}
\label{sec:intro}
Materials such as turbid water, white paint, and egg shells are considered opaque because multiple scattering by the randomly placed constituent scatterers in the medium frustrates the passage of light \cite{ishimaru1999wave}. The seminal  papers by Dorokhov \cite{dorokhov1982transmission}, Barnes and Pendry \etal \cite{pendry1990maximal,barnes1991multiple}, and others   \cite{mello1988macroscopic,beenakker2009applications} postulate that even if a normally incident wavefront barely propagates through a thick slab of such media, there will generically exist a few highly-transmitting wavefronts that will propagate through the slab  with a transmission coefficient close to $1$, \textit{i.e}, they will be nearly perfectly-transmitting. These perfectly transmitting eigen-wavefronts are the right singular vectors of the modal transmission matrix and are optimized to the specific random medium.

These seminal papers inspired the breakthrough experiments by Vellekoop and Mosk \cite{vellekoop2008phase,vellekoop2008universal}, and others \cite{popoff2010measuring,kohlgraf2010transmission,shi2010measuring,kim2012maximal,van2011optimal,aulbach2011control,cui2011high,cui2011parallel,stockbridge2012focusing}  provide credence to the hypothesis that there generally exist (nearly) perfectly transmitting eigen-wavefronts in highly scattering random media composed of a larger number of non-absorbing scatterers. Recently, we verified this hypothesis \cite{jin2012iterative,cjin2013} for 2-D systems with periodic boundary conditions composed of hundreds of thousands of non-absorbing scattering using numerically rigorous simulations.

The perfect-transmission supporting universal transmission coefficient distribution postulated by Dorokhov, Mello, Pereyra, Kumar, Pendry, and Barnes \cite{dorokhov1982transmission,pendry1990maximal,barnes1991multiple,mello1988macroscopic}, was derived assuming that the medium was deep enough so that the scattering matrix obeyed a physically consistent (\textit{i.e.} obeying reciprocity and time-reversal conditions) maximum-entropy law. Their analysis does not provide a principled and mathematically grounded framework for reasoning about whether, when or the sense in which a deep medium composed of a large number of randomly placed point-like scatterers with an arbitrary distribution of refractive indices can be expected to have a perfect transmission-supporting transmission coefficient distribution.

In this paper, we use modern random matrix theory to revisit the problem of predicting the transmission coefficient distribution of 2-D random media with periodic boundary conditions composed of a larger number of randomly placed point-like scatterers with an arbitrary refractive index distribution. We provide a characterization of the transmission coefficient distribution that explicitly depends on the refractive index distribution, the number of propagating modes and the depth of the medium for layered random media (in a sense we will make precise) composed of a large number of point-like scatterers.


The critical part of our derivation relies on the development of an isotropic random matrix model for the modal transfer matrix of a single randomly placed point-like scatterer. The random transfer matrix has  singular value distribution  that matches the singular values of the physical transfer matrix of a randomly-placed point-like scatterer. However, the left and right singular vectors of our random transfer matrix construction are modeled as independent and isotropically random. This allows us to use tools from free probability theory to approximate the transmission coefficient distribution of a layered random media composed of layers containing point-like scatterers.

We show that the derived distribution agrees remarkably well with results obtained using a numerically rigorous spectrally convergent simulation that utilizes spectrally accurate methodologies. This justifies the use of our isotropic model for reasoning about the properties of the derived distribution. Analysis of the resulting distribution brings into sharp focus the universal, \textit{i.e.}, scatterer-property independent, aspects of the distribution and provides the strongest principled justification yet of why we should expect perfect transmission in such deep random media regardless of the refractive index distribution of the constituent scatterers. The analysis brings into focus a sparsity condition under which random media can be expected to exhibit a perfect transmission-supporting universal transmission coefficient distribution in the deep medium limit.

We describe the setup and define the transmission coefficient distribution in Section \ref{sec:Setup} . We highlight some pertinent properties of the system modal transfer matrix in Section \ref{sec:transfer matrix}, and employ them in Section \ref{sec:transfer matrix model} to formulate a isotropically random model for the transfer matrix of a single point-like scatterer. In Section \ref{sec:tcoeff distribution}, we describe the pertinent free probabilistic tools from random matrix theory that allow us to analytically characterize the limiting transmission coefficient distribution of a medium composed of many scatterers from the eigen-distribution of the isotropic transfer matrix of a single point-like scatterer. We analyze the properties of the limiting transmission coefficient distribution thus obtained in Section \ref{sec:tcoeff properties}, and bring into sharp focus its universal, \textit{i.e.}, scatterer property independent, aspects. We validate our theoretical predictions using numerically rigorous simulations in Section \ref{sec:NumSim}. Details of some computations have been relegated to the Appendix.

\section{Setup}
\label{sec:Setup}

\begin{figure}[h]
\centering
\scalebox{0.5}{\input{ScatSysTCD5.pspdftex}}
\caption{Setup.}
\label{fig:setup}
\end{figure}

We study scattering from a two-dimensional (2D) random slab of thickness $L$ and periodicity $D$; the slab's unit cell occupies the space $0 \leq x < D$ and $0 \leq y < L$ (Fig.  \ref{fig:setup}). The slab contains $\Nlay$ infinite and $z$-invariant circular cylinders of radius $r$ that are placed randomly within the cell, as described shortly. The cylinders are assumed to be dielectric with refractive index $n_{d}$; care is taken to ensure the cylinders do not overlap. The radius of the cylinders is chosen to be much smaller than the wavelength $\lambda$ so they, in effect, act like point scatterers.

For $i_c = 1, 2, \ldots, \Nlay$, the $x$ and $y$ position of the center of the $i_c$-th cylinder are $u_{x,i_c}, u_{y,i_c} + (i_c -1) \ell)$, respectively where $u_{x,i_c}$ and $u_{y,i_c}$ s are i.i.d, random variables with uniform distribution on $[r,D-r]$  and $[r,\ell -r]$, respectively. Here $\ell= L/\Nlay$ is depth of each ``layer''; $\ell$ is chosen to be larger than $\sqrt{D \lambda}$. Each cylinder's refractive index $n_{i_c}$ is drawn independently from the same distribution of refractive indices $\eta(n)$.

Fields are $\sf TM_{\mbox{$z$}}$ polarized: electric fields in the $y<0$ $(i=1)$ and $y>L=\Nlay\,\ell$ $(i=2)$ halfspaces are denoted $\underline{e}_{i}(\underline{\rho})=e_{i}(\underline{\rho})\hat{z}$. The field amplitude $e_{i}(\underline{\rho})$ can be decomposed in terms of $+y$ and $-y$ propagating waves as $e_{i}(\underline{\rho}) = e_{i}^{+}(\underline{\rho}) + e_{i}^{-}(\underline{\rho})$, where
\begin{equation}
 \label{eq:IncidentWave}
e^{\pm}_{i}(\underline{\rho}) = \displaystyle \sum_{n=-N}^{N} h_{n} a^{\pm}_{i,n} e^{-j\underline{k}^{\pm}_{n} \cdot \underline{\rho}}\,.
\end{equation}
In the above expression, $\underline{\rho}=x\hat{x}+y\hat{y}\equiv(x,y)$, $\underline{k}^{\pm}_{n} = k_{n,x}\hat{x} \pm k_{n,y}\hat{y} \equiv (k_{n,x},\pm k_{n,y})$, $k_{n,x}=2\pi n/D$, $k_{n,y} = 2\pi \sqrt{(1/\lambda)^{2} - (n/D)^{2}}$, $\lambda$ is the wavelength, and $h_{n}=\sqrt{\| \underline{k}^{\pm}_{n} \|_{2} / k_{n,y}}$ is a power-normalizing coefficient; a time dependence $e^{j\omega t}$ is assumed and suppressed. We assume $N=\lfloor D/\lambda \rfloor$, \ie we only model propagating waves and denote $M=2N+1$. The modal coefficients $a^{\pm}_{i,n}$, $i=1,2$; $n=-N,\ldots,N$ are related by the scattering matrix
\begin{equation}\label{eq:scat matrix}
\left[\begin{array}{c}\aleftm\\\arightp\\\end{array}\right] = \underbrace{\left[ \begin{array}{cc} S_{11} & S_{12} \\ S_{21} & S_{22} \end{array} \right]}_{=:S} \left[\begin{array}{c}\aleftp\\\arightm\\\end{array}\right],
\end{equation}
where $\underline{a}^{\pm}_{i} = \begin{bmatrix} a^{\pm}_{i,-N} & \ldots a^{\pm}_{i,0} & \ldots a^{\pm}_{i,N}\end{bmatrix}^{T}$. In what follows, we assume that the slab is only excited from the $y<0$ halfspace; hence, $\arightm=0$. For a given incident field amplitude $e^{+}_{1}(\underline{\rho})$, we define the transmission coefficient as \begin{align}
\label{eq:tcoeff vec}
\tau(\aleftp) := \dfrac{\| S_{21}\cdot \aleftp \|_{2}^{2}}{\| \aleftp \|_{2}^{2}}.
\end{align}
 We denote the transmission coefficient of a normally incident wavefront by $\tnorm = \tau( \begin{bmatrix} 0 & \cdots & 1 &  \cdots &0 \end{bmatrix}^{T} )$; here $^{T}$ denotes transposition.

\subsection{The transmission coefficient distribution}\label{sec:tcoeff}

The problem of designing an incident wavefront $\aopt$ that maximizes the transmitted power can be stated as
\begin{equation}\label{eq:optimization problem}
\aopt= \argmax_{\aleftp}  \tau(\aleftp) = \argmax_{\aleftp} \dfrac{\| S_{21}\cdot \aleftp \|_{2}^{2}}{\| \aleftp \|_{2}^{2}} =  \argmax_{\parallel \aleftp\parallel_{2} = 1} \| S_{21} \cdot \aleftp \|_{2}^{2}
\end{equation}
where $\parallel \aleftp \parallel_{2} = 1$ represents an  incident power constraint.

Let $S_{21}= \sum_{i=1}^{M} \sigma_{i}\, \underline{u}_i \cdot \underline{v}_{i}^H$ denote the singular value decomposition (SVD) of $S_{21}$; $\sigma_{i}$ is the singular value associated with the left and right singular vectors $\underline{u}_i$ and $\underline{v}_i$, respectively. By convention, the singular values are arranged so that ${\sigma}_{1} \geq \ldots \geq {\sigma}_{M}$ and $^H$ denotes complex conjugate transpose. Then via a well-known result for the variational characterization of the largest right singular vector \cite[Theorem 7.3.10]{horn1990matrix}  we have that
\begin{equation}\label{eq:aopt s21}
\aopt = {\underline{v}}_{1}.
\end{equation}
When the optimal wavefront $\aopt$ is excited, the transmitted power is $ \topt :=\tau(\aopt) = \sigma_{1}^{2}$. When  the wavefront associated with the $i$-th right singular vector $\underline{v}_{i}$ is transmitted, the transmitted power is $\tau_{i} := \tau(\underline{v}_{i}) = \sigma_{i}^{2}$, which we refer to as the transmission coefficient of the $i$-th eigen-wavefront of $S_{21}$. We are interested in the limiting transmission coefficient distribution whose p.d.f. is defined as
\begin{equation}\label{eq:pdf tau}
f(\tau) = \lim_{M, \Nlay \to \infty} \mathbb{E} \left[\dfrac{1}{M} \sum_{i=1}^{M} \delta\left(\tau-\tau(\underline{v}_{i})\right) \right] = \lim_{M, \Nlay\to \infty} \mathbb{E}\left[\dfrac{1}{M} \sum_{i=1}^{M} \delta\left(\tau-\sigma_{i}^2)\right)\right],
\end{equation}
where we assume that $\Nlay/M \to c \in (0, \infty)$ as $M, \Nlay \to \infty$. The Dorokhov-Mello-Pereyra-Kumar (henceforth, DMPK)  distribution  \cite{dorokhov1982transmission,mello1988macroscopic} has density given by
\begin{equation}\label{eq:dmpk}
f_{\sf DMPK} (\tau) =  \dfrac{l_{\sf free}}{2L} \dfrac{1}{\tau \sqrt{1-\tau}}, \qquad \textrm{ for } 4 \exp(-L/2l_{\sf free}) \lessapprox \tau \leq 1,
\end{equation}
where $l_{\sf free}$ is the mean-free path in the medium. The DMPK distribution is posited \cite{dorokhov1982transmission,pendry1990maximal,barnes1991multiple,mello1988macroscopic,beenakker2009applications}
 to be the \textit{universal limiting distribution} for systems comprised of many scatterers in the limit where $L \gg M$.

Assuming a scattering regime where the DMPK distribution holds, Eq. (\ref{eq:dmpk}) predicts the existence of highly-transmitting eigen-wavefronts that achieve (nearly) perfect transmission.  Since the DMPK distribution was derived under a maximum-entropy type assumption  (which we shall revisit shortly), the material properties of the scatterers, such as the distribution of refractive indices, do not explicitly appear in the expression in Eq. (\ref{eq:dmpk}) for its p.d.f. but instead are encoded implicitly via the $l_{\sf free}$ parameter. Our objective is to theoretically predict $f(\tau)$ in Eq. (\ref{eq:pdf tau}) and explicitly characterize its dependence on the refractive index distribution $\eta(n)$, $\Nlay$, and $M$, assuming we are in a regime where each scatterer is small enough so that it effectively acts as an isotropic point scatterer. Our mathematically-derived framework permits reasoning about the conditions under which we might expect a universal limiting distribution and the existence of the (nearly) perfectly transmitting eigen-wavefronts.

\section{Background: the transfer matrix and its pertinent properties}\label{sec:transfer matrix}

The scattering matrix $S$ in Eq. (\ref{eq:scat matrix}) describes the relationship between the modal coefficients of incoming and outgoing waves. Rearranging the terms in Eq. (\ref{eq:scat matrix}) relates the modal coefficients in $i=1$ and $i=2$ halfspace via the transfer matrix $T$
\begin{equation}
{ \begin{bmatrix} \underline{a}_{2}^{+} \\  \\ \underline{a}_{2}^{-} \end{bmatrix} =  \underbrace{\begin{bmatrix} S_{21} - S_{22} \cdot S_{12}^{-1}\cdot S_{11} &   S_{22} \cdot S_{12}^{-1} \\   & \\ - {{S_{12}}}^{-1} \cdot S_{11} & S_{12}^{-1} \end{bmatrix}}_{=: \, T} \cdot \begin{bmatrix} \underline{a}_{1}^{+} \\ \\ \underline{a}_{1}^{-} \end{bmatrix}},
\end{equation}
where we have assumed that the $S_{12}$ matrix is invertible. Rewriting the transfer matrix as
\begin{equation}\label{eq:Tmatrix block}
 T = \begin{bmatrix} S_{22} & 0 \\ 0 & S_{12}^{-1} \end{bmatrix} \cdot \begin{bmatrix} S_{22}^{-1} \cdot S_{21} \cdot S_{11}^{-1} - S_{12}^{-1} & I \\
- I & S_{12} \end{bmatrix} \cdot \begin{bmatrix} S_{11} & 0 \\ 0 & S_{12}^{-1} \end{bmatrix},
\end{equation}
allows us to easily verify that $\det(T) = \det(T^{H}\cdot T) = 1$. In the lossless setting when $S^{H} \cdot S = I$, and $S_{12}$ is invertible, it is shown in Appendix \ref{sec:appendix transfer} that the $2M$ eigenvalues of $T^{H}\cdot T$ denoted by $\lambda_{1} \geq \ldots \geq \lambda_{2M}$ are
\begin{equation} \label{eq:tth eigs}  \lambda_{i} =  \dfrac{2 - \tau_i + 2 \sqrt{1 - \tau_i}}{\tau_i}\,\,\, \textrm{and}\,\, \lambda_{2M - i + 1} =  \dfrac{2 - \tau_i - 2 \sqrt{1 - \tau_i}}{\tau_i}  \qquad \textrm{ for } i = 1, \ldots, M.
\end{equation}
Note that $\lambda_{i} \cdot \lambda_{2M - i + 1} = 1$ so that the $2M$ eigenvalues of $T^{H}\cdot T$ come in reciprocal pairs. From Eq. (\ref{eq:tth eigs}), we have  that
$$ \lambda_{i} + \lambda_{2M - i + 1} = \dfrac{4}{\tau_i} - 2,$$
so that
\begin{equation}\label{eq:taui lambdai}
\tau_{i} = \dfrac{4}{\lambda_{i} + 1/\lambda_{i} + 2}.
\end{equation}
Substituting $\lambda_{i} = \exp(2 \,x_i)$ in Eq. (\ref{eq:taui lambdai}) yields
$$ \tau_i = \dfrac{4}{\exp(2 x_i) + \exp(-2 x_i) + 2} = \dfrac{1}{ (\exp(x_i) + \exp(-x_i)/2)^2} = \dfrac{1}{\cosh^2(x_i)}.
$$
Equivalently, since $x_{i} = 0.5 \, \ln \lambda_{i}$, we have
\begin{equation}\label{eq:conversion}
\tau_i = \dfrac{1}{\cosh^2(0.5 \ln \lambda_i)}  \leftrightarrow \lambda_i = \exp( 2 \cosh^{-1}(1/\sqrt{\tau_i})),
\end{equation}
and we have obtained a direct relationship between the eigenvalues of $T^H\cdot T$ and the transmission coefficients. Let $h(\lambda)$ denote the limiting eigenvalue distribution of the transfer matrix defined as
\begin{equation}\label{eq:pdf lambda}
h(\lambda) = \lim_{M, \Nlay \to \infty} \mathbb{E}\left[ \dfrac{1}{2M} \sum_{i=1}^{2M} \delta\left(\lambda-\lambda_i\right)\right].
\end{equation}
Then, a direct consequence of Eq. (\ref{eq:conversion}) is that once we know $h(\lambda)$, a simple change of variables yields the transmission coefficient distribution $f(\tau)$ as
\begin{align}\label{eq:ftau1}
  f(\tau) & = h(\lambda)  \dfrac{1}{{| \partial \tau}/{\partial \lambda |}} \bigg|_{\lambda \textrm{ in Eq. } (\ref{eq:conversion}) }  = \,\, h(\lambda)  \dfrac{(\lambda+1)^{3}}{4 |\lambda-1|} \bigg|_{\lambda = \exp( 2 \cosh^{-1}(1/\sqrt{\tau}))}.
\end{align}
Since the eigenvalues of $T^{H}\cdot T$ come in reciprocal pairs, $h(\lambda)$ for $\lambda \leq 1$ uniquely determines $h(\lambda)$ for $\lambda >1$. Thus we can rewrite Eq. (\ref{eq:ftau1}) as
\begin{equation}\label{eq:ftau2}
    f(\tau) =  2 h(\lambda) \, \mathbb{I}_{\lambda \leq 1}  \dfrac{(\lambda+1)^{3}}{4 | \lambda-1 |} \bigg|_{\lambda = \exp( 2 \cosh^{-1}(1/\sqrt{\tau}))},
\end{equation}
where $\mathbb{I}_{\lambda \leq 1}$ denotes the indicator function on the set $\lambda \leq 1$. From Eq. (\ref{eq:taui lambdai}), we have that
$$ \dfrac{1}{\tau} = \dfrac{(\lambda+1)^{2}}{4\,\lambda} \quad \textrm{ and } \qquad \dfrac{1}{{1-\tau}} = \dfrac{(\lambda+1)^{2}}{(\lambda-1)^{2}},
 $$
 so that rearranging terms on the right hand side of Eq. (\ref{eq:ftau2}), yields
 \begin{align}\label{eq:ftau conversion}
  f(\tau) & =\dfrac{1}{\tau \sqrt{1- \tau}} \cdot \{ 2 \, h(\lambda)  \lambda \, \mathbb{I}_{\lambda \leq 1}\}\bigg|_{\lambda = \exp( 2 \cosh^{-1}(1/\sqrt{\tau}))}.
 \end{align}
We note that Eq. (\ref{eq:ftau conversion}) is an exact relationship between the eigenvalue distribution of the transfer matrix and the transmission coefficient distribution. Comparing Eqs. (\ref{eq:ftau conversion}) and (\ref{eq:dmpk}) reveals the important insight that the DMPK distribution arises under the assumption that in the limit of deep random media, $h(\lambda) = l_{\sf free}/(4 L \lambda) $, or equivalently, that $h(\lambda)$ is a \textit{log-uniform distribution}. This is the \textit{maximum-entropy assumption} that yields the DMPK distribution for deep random media. Our goal is to analytically characterize $h(\lambda)$ and hence $f(\tau)$, via Eq. (\ref{eq:ftau conversion}) as a function of $\Nlay$, $M$ and the refractive index distribution of the scatterers for the setup in Fig. \ref{fig:setup}.

\section{An isotropically random model for the transfer matrix of a single point-like scatterer}\label{sec:transfer matrix model}
Let $T_{i}$ denote the transfer matrix of a layer containing a single scatterer (Fig. \ref{fig:setup}) and let $S^{(i)}$, $S_{11}^{(i)}$, $S_{22}^{(i)}$, and $S_{21}^{(i)}$ denote its scattering matrix and subblocks thereof, respectively. When the scatterers are point-like and $D$ is large, then $S_{11}$ and $S_{22}$ are well approximated by a rank one matrix whose largest singular value $\alpha\in[0,1)$ we will refer to as the scattering strength. This is obviously true for $D \to \infty$, and remains remarkably accurate for smaller $D$ as well. Since $S$ is unitary for lossless media, we have that $S_{11}^{H}\cdot S_{11} + S_{21}^{H} \cdot S_{21} = I$. Hence, the $S_{21}$ matrix must have an SVD of the form
\begin{align*}
  S_{21}^{(i)} = U_i \cdot \Diag (1,\ldots, 1, \sqrt{1-\alpha^{2}}) \cdot V_i^{H},
\end{align*}
where $U_{i}$ and $V_{i}$ are the left and right singular vectors of $S_{21}^{(i)}$, which encode the physics of the scattering system. Consequently, by Eq. (\ref{eq:tcoeff vec}), the transmission coefficients of $S_{21}^{(i)}$ are approximately
$$\tau_{1} \approxeq \ldots \approxeq \tau_{M-1} \approxeq 1,  \textrm{ and } \tau_{M}  \approxeq 1-\alpha^{2}.$$
From Eq. (\ref{eq:tth eigs}), we can conclude that the $2M-2$ eigenvalues of $T^{H}\cdot T$ will equal one. The remaining two eigenvalues will $\lambda_{M}$ and $\lambda_{M+1} = 1/\lambda_{M}$ where
\begin{equation}\label{eq:theta alpha}
\lambda_{M} = \dfrac{2- (1-\alpha^{2}) + 2 \sqrt{1-(1-\alpha^2)}}{1-\alpha^{2}} = \dfrac{1+ \alpha^{2} + 2\,\alpha}{(1-\alpha)(1+\alpha)} = \dfrac{1+\alpha}{1- \alpha}=: \theta.
\end{equation}
This implies that the transfer matrix will have an SVD of the form
\begin{equation}\label{eq:transfer matrix svd}
T_{i} =  \widetilde{U}_{i} \, {\rm  diag}(\underbrace{1, 1, \cdots, 1, 1}_{2M - 2 \, \textrm{entries}}, \sqrt{\theta}, 1/\sqrt{\theta}) \, \widetilde{V}_{i}^{H},
\end{equation}
where $\widetilde{U}_{i}$ and $\widetilde{V}_{i}$ are the left and right singular vectors of $T_{i}$, which again encode the physics of the scattering systems. The refractive index distribution $\eta(n)$ induces a distribution $f_\theta(t)$ on $\theta$ which we assume to known and obtained either a via a change of variables as
\begin{equation}\label{eq:point change of variables}
 \alpha \approx 9 \sqrt{\dfrac{ r^{4} (2\pi/\lambda)^{3} \pi^2 (n_d^2 - 1)^2}{16 D}},
\end{equation}
under a point scatterer assumption for the large $D$, $r \ll \lambda$, and $n_d \approx 1 $ regime, or using computational electromagnetic techniques.

\begin{figure}[H]
\centering
\label{fig:incoherence}
\subfloat[3D plot.]{
\includegraphics[width=0.5\textwidth]{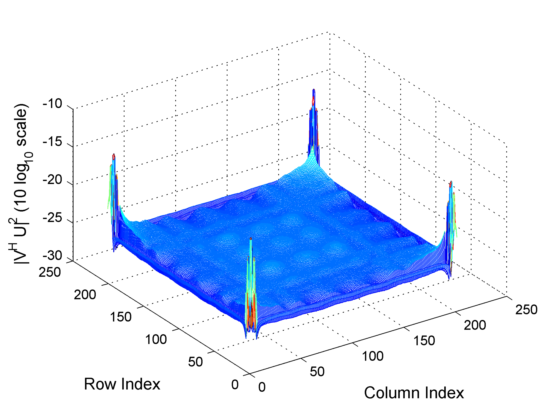}
}
\subfloat[Top view.]{
\includegraphics[width=0.5\textwidth]{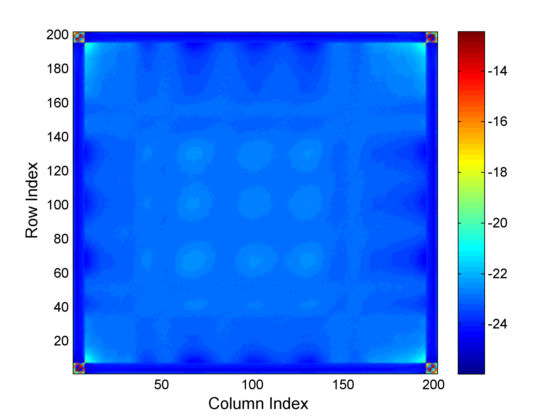}
}
\caption{Relationship between singular vectors plotted in $10\log_{10}$ scale. The absolute value squared of the correlation matrix between the singular vectors averaged over 100000 trials. The settings were $n=3.1, \alpha=0.9, \theta = 0.053, r = 0.05\lambda, \ell = 11.66\lambda, D = 50.43\lambda, M=101$.}
\end{figure}

The transfer matrix of the entire system in Fig. \ref{fig:setup} is obtained from those of the layers as
\begin{equation}\label{eq:transfer product}
T = \prod_{i=1}^{\Nlay} T_{i}.
\end{equation}
Each of the transfer matrices are independent and identically distributed. Fig. \ref{fig:incoherence} plots the expected values of the squared magnitude of the (bistochastic) correlation matrix formed by the inner product of the right singular vectors of a transfer matrix associated with a single randomly placed scatterer and the left singular vectors of an independent transfer matrix associated with another randomly placed scatter, averaged over $100,000$ independent realizations. If the singular vectors were independent and isotropically random (or Haar distributed) then we would get an empirically averaged matrix with all of its entries close to $1/M$. From Fig. \ref{fig:incoherence}, we can conclude that the singular vectors of two independent transfer matrices are not isotropically random with respect to each other. However, most of the entries of the correlation matrix have entries  `close' to $1/M$.

Very recently, Anderson and Farrell \cite{anderson2014asymptotically}  rigorously showed that the product of independent (Hermitian) random matrices with independent eigenvectors having a correlation matrix whose entries have squared magnitude entries exactly equal to $1/M$ will have the same limiting distribution as the product of independent random matrices with the same eigenvalue distribution but isotropically random eigenvectors. Here too, we have a situation where we are interested in analyzing the singular value distribution of products of random matrices with independent singular vectors. However, the correlation matrix of the singular vectors has entries whose squared magnitude is not exactly $1/M$, as would be the case if the left and right singular vectors were isotropically random, but instead \textit{close to} $1/M$. This leads to our conjecture that the singular value distribution of the matrix in Eq. (\ref{eq:transfer product}) can be `well approximated by' the singular value distribution of independent random matrices with the same per-matrix singular value distribution but isotropically random left and right singular vectors.

Motivated by this conjecture, we now consider an isotropic random matrix model for the transfer matrices whose singular values are specified by Eq. (\ref{eq:transfer matrix svd}) but whose left and right singular vectors are independent and isotropically random. We then use tools from free probability theory to analytically characterize the transmission coefficient distribution that arises due to this isotropic model for the transfer matrix of a point-like scatterer. Numerically rigorous physical simulations in Section \ref{sec:NumSim} will validate our conjecture. A mathematically rigorous treatment of this conjecture, including a quantification of the approximation error, remains an open problem.

\section{Analytically characterizing the transmission coefficient distribution}\label{sec:tcoeff distribution}
We now discuss some preliminaries required to compute the transmission coefficient distribution under the isotropic transfer matrix assumption. Let  $X_M$ be  an $M \times M$ symmetric (or Hermitian) random matrix whose ordered eigenvalues are denoted by $\lambda_{1}(X_M) \geq \cdots \geq \lambda_{n}(X_M)$. Let $h_{X_{M}}$ be the empirical eigenvalue distribution, \ie, the probability distribution with density
\[
h_{X_{M}}(z) = \frac{1}{M} \sum_{i=1}^{M} \delta\left(z-\lambda_{i}(X_M)\right).
\]
Now suppose that $A_M$ and $B_M$ are two independent $M \times M$ matrices whose empirical eigenvalue distributions converge as $M \longrightarrow \infty$ to non-random distributions having densities $h_{A}$ and $h_{B}$, respectively. A natural question then is: how is the limiting eigenvalue distribution of the matrix $B_{M}^{H} \cdot A_{M}^{H} \cdot A_M \cdot B_M $ related to the limiting eigenvalue distributions of $A_{M}$ and $B_{M}$?

Free probability theory \cite{voiculescu1991limit,dv1992free} states  that if we know $h_A$ and $h_B$ and the matrices $A_M$ and $B_M$ are \textit{asymptotically free}, we can compute the limiting eigenvalue distribution of $A_M \cdot B_M$ from the limiting eigenvalue distributions of $A$ and $B$. Specifically, in this setting, $h_{A \cdot B}$ is given by the free multiplicative convolution of $h_A$ and $h_B$, denoted by $h_{A} \boxtimes h_{B}$ which  is computed as described next. We first define the {\it $S$-transform}\footnote{Denoted here by $\psi(\cdot)$ to avoid any confusion with the $S$ (or scattering) matrix.},  which is given by
\begin{equation}\label{eq:S transform}
\psi_X(z):= \dfrac{1+z}{z } \cdot \dfrac{1}{\xi_X^{-1}(z)},
\end{equation}
where
\begin{equation}\label{eq:xiz defn}
\xi_X(z) =\int\dfrac{t}{z-t} h_X(t) dt =-1 + z \,g_X(z),
\end{equation}
and
\begin{equation}
g_{X} (z) = \int\dfrac{1}{z-t} h_X(t) dt,
\end{equation}
is the Cauchy transform of $h_X$. Then $S$-transform of $h_{A\cdot B}$ is
\begin{equation}
\psi_{A\cdot B}(z)= \psi_{A}(z)\psi_{B}(z).
\end{equation}
Note, that given the Cauchy transform $g_X(z)$, we can recover the density via the inversion formula
\begin{equation}\label{eq:cauchy inversion}
 h_X(z) = - \dfrac{1}{\pi} \lim_{\epsilon \to 0} \Imag \, g_{X}(z+j \,\epsilon).
\end{equation}

A sufficient condition for the asymptotic freeness of two random matrices is that their singular vectors are independent and isotropically random \cite{hiai2000semicircle}. Consequently, under the isotropic transfer matrix assumption, the transfer matrices of successive layers are asymptotically free, by construction. Hence, we can use free multiplicative convolution machinery to characterize the limiting eigenvalue distribution of the transfer matrix of a multi-layered scattering system as depicted in Fig. \ref{fig:setup}, since, by Eq. (\ref{eq:transfer product}), the composite transfer matrix is the product of $\Nlay$ independent (and asymptotically free) random transfer matrices each having independent, isotropically random left and right singular vectors and singular values given by Eq. (\ref{eq:transfer matrix svd}).

To that end, we first compute the empirical eigenvalue distribution of $T_{i}^{H} \cdot T_{i}$ which is
\begin{equation}\label{eq:h kequals1}
h_{i}(\lambda) = \left(1-\dfrac{2}{2M}\right) \delta(\lambda-1)  + \dfrac{1}{2M} \delta(\lambda- \theta)  +  \dfrac{1}{2M} \delta(\lambda- 1/\theta).\end{equation}
Its Cauchy transform is given by
\begin{equation}
g_{i}(z) = \left(1-\dfrac{1}{M}\right) \dfrac{1}{z-1}  + \dfrac{1}{2M} \left( \dfrac{1}{z-\theta}+  \dfrac{1}{z-1/\theta}\right),
\end{equation}
and
\begin{align}\label{eq:xiz}
\begin{split}
\xi_{i}(z) &= -1 + \left(1-\dfrac{1}{M}\right) \dfrac{z}{z-1}  + \dfrac{1}{2M} \left( \dfrac{z}{z-\theta} +  \dfrac{z}{z-1/\theta}\right)\\
& = -1 + \dfrac{z}{z-1} - \dfrac{1}{M} \left(\dfrac{z}{z-1} - \dfrac{0.5 \, z}{z-\theta} -  \dfrac{0.5\,z}{z-1/\theta} \right)\\
& = \underbrace{\dfrac{1}{z-1}}_{=: \,\xi_{0}(z)} + \dfrac{1}{M} \underbrace{\left( \dfrac{0.5 \, z}{z-\theta} +  \dfrac{0.5\,z}{z-1/\theta} -\dfrac{z}{z-1}  \right)}_{=: \,\widetilde{\xi}(z)}.
\end{split}
\end{align}
Repeating the computation for the setting where the $\theta_i$'s are random with pdf $f_{\theta}(\cdot)$ yields
\begin{equation}\label{eq:xi general}
\widetilde{\xi}(z)= \displaystyle z \int \left[ \dfrac{0.5}{z-t} +  \dfrac{0.5}{z-1/t} -\dfrac{1}{z-1}  \right] f_{\theta}(t) dt .
\end{equation}

To compute $\psi_{i}(z)$ using Eq. (\ref{eq:S transform}) we need to compute $\xi_{i}^{-1}(z)$. A standard application of perturbation theory (see, e.g., \cite{schwartz1977classical}) yields
\begin{equation}\label{eq:finversion}
  \xi_i^{-1}(z) = \xi_{0}^{-1}(z)  - \dfrac{1}{M} \, \dfrac{\widetilde{\xi}(x)}{\partial_{x} \xi_{0}(x)} \bigg|_{x = \xi_{0}^{-1}(z)} + O\left(\dfrac{1}{M^{2}}\right).
 \end{equation}
Substituting $\xi_{0}^{-1}(z) = (z+1)/z$ gives
\begin{equation}\label{eq:finversion}
  \xi_i^{-1}(z) = \dfrac{z+1}{z} - \dfrac{1}{M} \, \dfrac{\widetilde{\xi}(x)}{\partial_{x} \xi_{0}(x)} \bigg|_{x = 1+1/z} + O\left(\dfrac{1}{M^{2}}\right),
 \end{equation}
 or equivalently
 $$ \dfrac{z}{z+1} \, \xi_i^{-1}(z) = 1 +\dfrac{1}{M}  \dfrac{z}{z+1}  \cdot \dfrac{\widetilde{\xi}(1+\tfrac{1}{z})}{z^2}
+  O\left(\dfrac{1}{M^{2}}\right),$$
so that by Eq. (\ref{eq:S transform}),
$$ \psi_{i}(z) = 1 - \dfrac{1}{M}  \dfrac{\widetilde{\xi}(1+\tfrac{1}{z}) }{z(z+1)}
+  O\left(\dfrac{1}{M^{2}}\right).
$$
Then,
\begin{align}
  \psi_{h}(z) & = \prod_{i=1}^{\Nlay} \psi_{i}(z) =\left[ 1 - \dfrac{1}{M}  \dfrac{\widetilde{\xi}(1+\tfrac{1}{z}) }{z(z+1)}
+  O\left(\dfrac{1}{M^{2}}\right) \right]^{\Nlay}
\end{align}

In the regime where $M, \Nlay \to \infty$ with $\Nlay/M \to c$ we obtain
\begin{align}\label{eq:psi final}
\begin{split}
  \psi_{h}(z; c) & = \lim_{M \to \infty}  \left[ 1 - \dfrac{1}{M}  \dfrac{\widetilde{\xi}(1+\tfrac{1}{z}) }{z(z+1)}
+  O\left(\dfrac{1}{M^{2}}\right) \right]^{M \cdot \frac{\Nlay}{M}}\\
  & = \left[\exp\left( - \dfrac{\widetilde{\xi}(1+\tfrac{1}{z}) }{z(z+1)}\right)\right]^{c} = \exp\left( -c \cdot \dfrac{\widetilde{\xi}(1+\tfrac{1}{z}) }{z(z+1)}\right).
\end{split}
\end{align}

We next discuss how to obtain the distribution from $\psi_{h}(z;c)$. Inserting $z=\xi_h(y)$ in Eq. (\ref{eq:S transform}), we get
\begin{equation}
\dfrac{1+\xi_h(y)}{\xi_h(y) } \dfrac{1}{y} = \psi_h(\xi_{h}(y)).
\end{equation}
Substituting in the expression for $\xi_{h}(y)$ from Eq. (\ref{eq:xiz defn}), we get
\begin{equation}
\dfrac{1-1 + y \,g_h(y)}{-1 + y \,g_h(y) } \dfrac{1}{y} = \psi_h(-1 + y \,g_h(y)).
\end{equation}
Therefore, we get the fixed-point equation
$$\dfrac{g_{h}(z)}{z g_{h}(z) - 1} = \psi_h(zg_{h}(z)-1).$$
Substituting Eq. (\ref{eq:psi final})
\begin{equation}
\label{eq:fixedptEq}
\dfrac{g_{h}(z)}{z g_{h}(z) - 1}  = \exp\left[ -c \cdot \dfrac{\widetilde{\xi}(\frac{z g_h(z)}{z\,g_h(z) -1} ) }{z g_h(z)\,(z \,g_h(z)+1)}\right]
\end{equation}
The density $h(\lambda)$ can be recovered from the Cauchy transform $g_h(\lambda)$ using Eq. (\ref{eq:cauchy inversion}) after solving the fixed-point equation. The transmission coefficient distribution is then obtained by Eq. (\ref{eq:ftau conversion}). Note that $\widetilde{\xi}(z)$ in Eq. (\ref{eq:xi general}) explicitly encodes the portion of the limiting distribution that depends on the scatterer-dependent properties via $f_\theta(t)$, where $\theta$ is related to the scattering strength $\alpha$ of a single scatterer via Eq. (\ref{eq:theta alpha}) and $\alpha$ is related to the scatterer-dependent properties via Eq. (\ref{eq:point change of variables}).

\section{Properties of the limiting transmission coefficient distribution}\label{sec:tcoeff properties}
We now analyze the properties of the distributions characterized by Eq. (\ref{eq:fixedptEq}). The mean of $f(\tau)$ is
\begin{equation}\label{eq:mean}
 \mathbb{E}[\tau] = \int \tau f(\tau) d\tau = \displaystyle \int \dfrac{4 }{\lambda + \lambda^{-1} + 2} h(\lambda) d\lambda  = 4 \displaystyle\int \dfrac{\lambda}{(\lambda+1)^{2}} h(\lambda) d\lambda
 \end{equation}
where we have used Eq. (\ref{eq:taui lambdai}) to express $\mathbb{E}[\tau]$ with respect to $h(\lambda)$. From Eq. (\ref{eq:xiz defn}), we note that
\begin{equation}\label{eq:dxiz}
\xi_{h}'(z) := \partial_{z} \xi_{h}(z) = - \displaystyle\int \dfrac{\lambda}{(z-\lambda)^{2}} h(\lambda) d\lambda.
\end{equation}
Thus by comparing the righthand sides of Eqs. (\ref{eq:mean}) and (\ref{eq:dxiz}), we have that
\begin{equation}\label{eq:mean2}
 \mathbb{E}[\tau] = -4\, \xi_{h}'(-1).
 \end{equation}

From the computation in Appendix \ref{sec:first moment}, we obtain the closed-form expression
 \begin{align}
 \begin{split}\label{eq:first moment closed form}
 \mathbb{E}[\tau] = \dfrac{1}{1+c\, \underbrace{\displaystyle\int \left( \dfrac{1- t}{1 + t} \right)^2 f_{\theta}(t) dt}}_{=:B_{2}} = \dfrac{1}{1+cB_{2}}.
 \end{split}
\end{align}
Here, we call $B_{2}$ the normalizing factor, and it represents the average scattering strength of a single layer. The normalizing factor can be used to homogenize two different materials by giving measures to calculate the effective lengths, and its specific usage will be discussed in Section \ref{sec:NumSim}. We now compute the second moment of $f(\tau)$, which is given by
\begin{equation}\label{eq:second moment}
 \mathbb{E}[\tau^{2}] = \int \tau^{2} f(\tau) d\tau = \displaystyle \int \dfrac{16 }{(\lambda + \lambda^{-1} + 2)^{2}} h(\lambda) d\lambda  = 16 \displaystyle\int \dfrac{\lambda^{2}}{(\lambda+1)^{4}} h(\lambda) d\lambda .
 \end{equation}
Note that
$$ \xi_{h}''(z) := \partial_{z} \xi_{h}'(z) = -2 \displaystyle\int \dfrac{\lambda}{(\lambda-z)^{3}} h(\lambda) d\lambda,$$
and
$$\xi_{h}'''(z)  := \partial_{z} \xi_{h}''(z) = -6 \displaystyle \int \dfrac{\lambda}{(\lambda-z)^{4}} h(\lambda) d\lambda,$$
so that
\begin{equation}\label{eq:xi second moment}
  \dfrac{1}{6}  \xi_{h}'''(z) - \dfrac{1}{2} \xi_{h}''(z)  = \displaystyle\int \left[- \dfrac{\lambda}{(\lambda-z)^{4}}+  \dfrac{\lambda}{(\lambda-z)^{3}} \right] h(\lambda) d\lambda = \displaystyle\int  \dfrac{\lambda^{2} - \lambda\,z- \lambda}{(\lambda-z)^{4}} h(\lambda) d\lambda.
 \end{equation}
 Comparing Eqs. (\ref{eq:second moment}) and (\ref{eq:xi second moment}) gives us the relationship
 \begin{equation}\label{eq:second moment}
 \mathbb{E}[\tau^{2}] = 16 \left[ \dfrac{1}{6}  \xi_{h}'''(-1) - \dfrac{1}{2} \xi_{h}''(-1)  \right].
 \end{equation}

The closed-from expression for the second moment is lengthy and  derived in Appendix \ref{sec:second moment}.
From Eq. (\ref{eq:Etau sq calc}) and Eq. (\ref{eq:mean3})  we obtain
\begin{subequations}
\begin{align}
  \dfrac{\mathbb{E}[\tau^2]}{\mathbb{E}[\tau]} &= \dfrac{16}{-4} \dfrac{\dfrac{1}{6}  \xi_{h}'''(-1) - \dfrac{1}{2} \xi_{h}''(-1) }{\xi_{h}'(-1)}\\
  &= -4 \dfrac{\dfrac{1}{6} \dfrac{3\left( \xi_{h}^{-1''} \right)^2 - \xi_{h}^{-1'} \xi_{h}^{-1'''}}{\left(\xi_{h}^{-1'}\right)^5} + \dfrac{1}{2} \dfrac{\xi_{h}^{-1''}}{\left( \xi_{h}^{-1'} \right)^3}  }{ \dfrac{1}{\xi_{h}^{-1'} }}\\
  \label{eq:Ratio}&= \dfrac{-2}{3} \dfrac{ 3\left( \xi_{h}^{-1''} \right)^2 - \xi_{h}^{-1'} \xi_{h}^{-1'''} + 3(\xi_{h}^{-1'})^2 \xi_{h}^{-1''}}{ (\xi_{h}^{-1'})^{4} }.
\end{align}
\end{subequations}

The exact (cumbersome) expression for the ratio can be obtained by plugging in Eqs. (\ref{eq:xiInv1new}), (\ref{eq:xiInv2}) and (\ref{eq:xiInv3}) into Eq. (\ref{eq:Ratio}).

\subsection{Universal aspects of the limiting distribution}
We now consider the $c \rightarrow \infty$ properties of the limiting distribution. Consider the ratio $\mathbb{E}[\tau^2]/\mathbb{E}[\tau]$. 
To that end, we isolate the highest order term of $c$ in the denominator and numerator and obtain
\begin{align}
  \dfrac{\mathbb{E}[\tau^2]}{\mathbb{E}[\tau]} = \dfrac{2}{3} \dfrac{2^{12} c^4 B_8 + O(c^3)}{2^{12} c^4 B_8 +O(c^3)}.
\end{align}
We arrived at this expression by manipulating the expressions for $\mathbb{E}[\tau]$ and $\mathbb{E}[\tau^2]$ given by Eq. (\ref{eq:mean3}) and Eq. (\ref{eq:second moment2}) respectively, that involved the terms in Eqs. (\ref{eq:xiInv1new})- (\ref{eq:xiInv3}). Therefore, \begin{equation}\label{eq:limiting ratio}
  \lim_{c \rightarrow \infty} \dfrac{\mathbb{E}[\tau^2]}{\mathbb{E}[\tau]} = \dfrac{2}{3}.
\end{equation}
This limiting ratio is universal in the sense that it does not depend on $f_{\theta}$ and coincides with the answer obtained by integrating the DMPK distribution \cite{mello1988macroscopic,beenakker1997random,vellekoop2008universal}

We will now compute the first two moments of the DMPK distribution in  Eq. (\ref{eq:dmpk}). Let us suppose that the eigenvalues of the transfer matrix are log-uniformly distributed so that $h(\lambda) \lambda = \kappa \mathcal{I}_{[\epsilon,1/\epsilon]}$ for some small positive $\epsilon$ such that $\epsilon \ll 1$. Then Eq. (\ref{eq:mean}) gives us
$$ \mathbb{E}[\tau] = 4 \,\kappa \,\displaystyle\int_{0}^{\infty} \dfrac{1}{(\lambda+1)^{2}} d\lambda + O(\epsilon)= 4 \kappa + O(\epsilon),$$
whereas Eq. (\ref{eq:second moment}) gives us
$$ \mathbb{E}[\tau^{2}] = 16 \,\kappa \,\displaystyle\int_{0}^{\infty} \dfrac{\lambda}{(\lambda+1)^{4}} d\lambda  + O(\epsilon) = \dfrac{16}{6} \kappa + O(\epsilon),$$
so that
$$
\dfrac{\mathbb{E}[\tau^2]}{\mathbb{E}[\tau]} = \dfrac{16/6}{4} = \dfrac{2}{3} + O(\epsilon),$$
and for $\epsilon \ll 1$, we get the universal limiting ratio  predicted Eq. (\ref{eq:limiting ratio}). Thus in the $c \to \infty$ limit the DMPK distribution exhibits the same universal ratio of the first and second moments as the limiting distribution we have derived using random matrix theoretic arguments.

The discussion in Section \ref{sec:k points} suggests that whenever the medium is `sparse' in the sense that $k_M \, \Nlay/M \to c$, we can expect to get a distribution of the form posited by the DMPK theory irrespective of the material properties of the individual scatterers. The natural next step in this line of inquiry is to analyze the large $c$ asymptotics of the transmission coefficient distribution via its implicit characterization in Eq. (\ref{eq:fixedptEq}) to answer finer questions about the existence of a density at $\lambda = 1$ (equivalently $\tau = 1$) for all $ c \in (0, \infty)$. We leave these for future work.

\subsection{Multiple-point-scatterer-per-layer scenarios that lead to same limiting distribution}\label{sec:k points}
We now consider multiple-point-scatterer-per-layer scenarios that lead to the same limiting distribution - this will suggest a sparsity condition for the existence of the perfect transmission-supporting universal limiting transmission coefficient distribution. Consider the setting similar to that in Fig. \ref{fig:setup} except with $k$ randomly placed point scatterers per layer. Then if $D \gg r$ is large, we expect the $S_{11}$ and $S_{22}$ matrix to be approximately rank $k$, by neglecting the scatterer-scatterer interaction related terms. Consequently, we can model the empirical eigenvalue distribution of $T_i^{H}\cdot T_{i}$ as
\begin{equation}\label{eq:h general k}
h_{i}(\lambda) = \left(1-\dfrac{2k}{2M}\right) \delta(\lambda-1)  + \dfrac{k}{2M} \sum_{j=1}^{k}\left[\delta(\lambda- \theta_j)  +  \delta(\lambda- 1/\theta_j) \right].\end{equation}
Retracing the steps after Eq. (\ref{eq:h kequals1}), we observe that we arrive at the same limiting distribution encoded in Eq. (\ref{eq:fixedptEq}) except now with $k \, \Nlay/M \to c$ and
$$f_{\theta}(t) := \dfrac{1}{k} \sum_{j=1}^{k} f(\theta_j).$$

Now, suppose that we are in the setting where the rank of the $S_{11}$ and $S_{22}$ matrices depends on $M$. Let us make this dependence explicit by denoting it as $k_M$. Suppose that $k_M/M \to 0$. Then, following the argument following Eq. (\ref{eq:h general k}), we will arrive at the same limiting distribution encoded in Eq. (\ref{eq:fixedptEq}) whenever $k_M \, \Nlay/M \to c$ and
$$f_{\theta}(t) := \dfrac{1}{k_M} \sum_{j=1}^{k_M} f(\theta_j).$$

Our analysis thus suggests the sparsity condition $k_M/M \to 0$ and  $k_M \, \Nlay/M \to c$ for the emergence of the perfect transmission-supporting universal transmission coefficient distribution.

\section{Numerical simulations}
\label{sec:NumSim}
To validate the predicted transmission coefficient distribution, we adopt the numerical simulation protocol described in \cite{cjin2013}. Specifically, we  compute the scattering matrices in Eq. (\ref{eq:scat matrix}) via a spectrally accurate, T-matrix inspired integral equation solver that characterizes fields scattered from each cylinder in terms of their traces expanded in series of azimuthal harmonics. As in \cite{cjin2013}, interactions between cylinders are modeled using 2D periodic Green's functions. The method constitutes a generalization of that in \cite{mcphedran1999calculation}, in that it does not force cylinders in a unit cell to reside on a line but allows them to be freely distributed throughout the cell.  As in \cite{cjin2013}, all periodic Green's functions/lattice sums are rapidly evaluated using a recursive Shank's transform using the methods described in \cite{singh1991use,sidi2003practical}.  Our method exhibits exponential convergence in the number of azimuthal harmonics used in the description of the field scattered by each cylinder.  As in \cite{cjin2013}, in the numerical experiments below, care was taken to ensure 11-digit accuracy in the entries of the computed scattering matrices.

We now describe how the simulations were performed. We generated a random scattering system with $r = 0.05\lambda, \ell = 12.63\lambda, D = 25.75\lambda$, and $M=51$. The locations of the scatterers were selected randomly as described in Section \ref{sec:Setup}. For a given $\Nlay$, the number of layers in the scattering system, we numerically compute the scattering matrices. We then compute the empirical transmission coefficient distribution over $200$ Monte-Carlo trials and compare it to the analytically predicted transmission coefficient distribution obtained as a fixed point of Eq. (\ref{eq:fixedptEq}) for $c = \Nlay/M$ and an appropriate choice of $f_{\theta}$.

\begin{figure}
\centering
\subfloat[C=2.]{
\includegraphics[trim = 22 0 30 20, clip=true,width=0.45\textwidth]{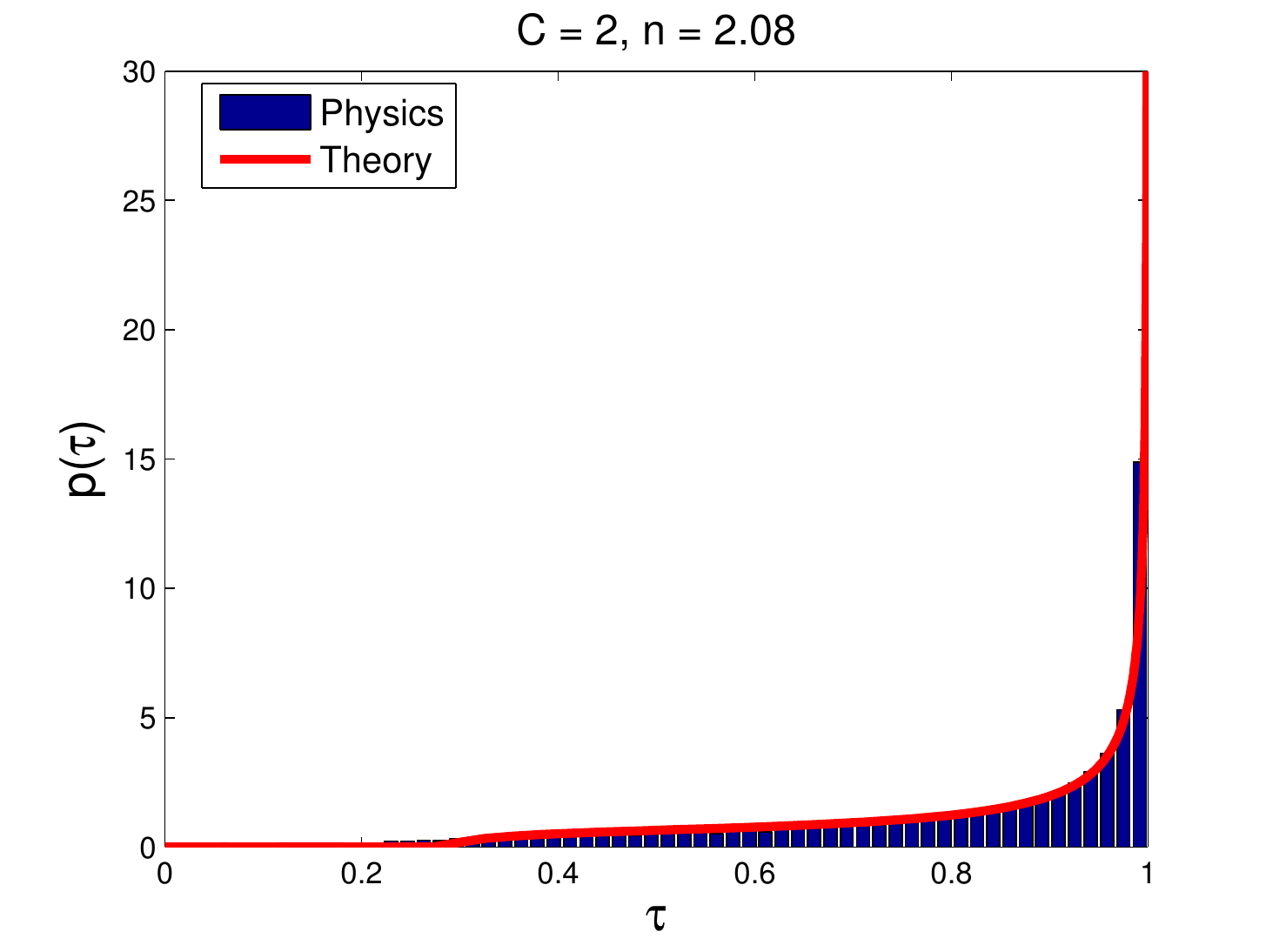}
}
\subfloat[C=41.]{
\includegraphics[trim = 22 0 30 20, clip=true,width=0.45\textwidth]{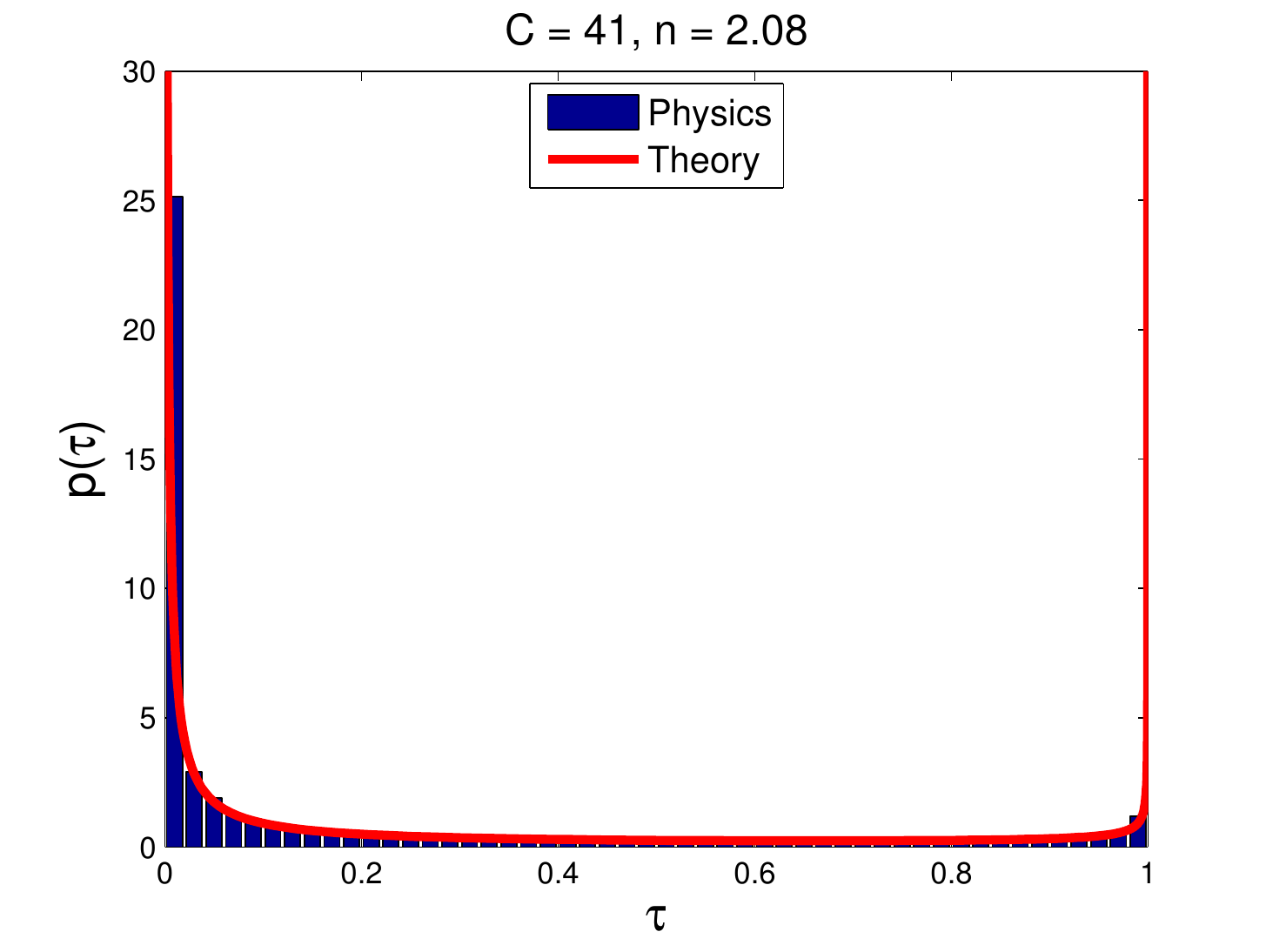}
}\\
\subfloat[C=2: Zoom-in.]{
\includegraphics[trim = 130 0 20 150, clip=true,width=0.70\textwidth]{PhysicsTheoryNonRandK1C2.pdf}
}\\
\subfloat[C=41: Zoom-in.]{
\includegraphics[trim = 40 0 20 200, clip=true,width=0.70\textwidth]{PhysicsTheoryNonRandK1C41.pdf}
}

\caption{The transmission coefficient distribution for the setting where $f_{\theta}(t) = \delta(t - \theta)$. The red line is the theoretical prediction - the histograms are from the physically rigorous simulation averaged over $100$ trials. Note the agreement with theory in the $C=2$ where the distribution is far from the DMPK distribution.}
\label{fig:nonrand}
\end{figure}

We first consider the setting where all the randomly placed cylinders have the same refractive index. Plugging in  $f_{\theta}(t) = \delta(t - \theta)$ into Eq. (\ref{eq:xi general}) yields the expression
\begin{equation}\label{eq:xi nonrandom}
\widetilde{\xi}(z) = z\left( \dfrac{1}{z-\theta} + \dfrac{1}{z-\theta^{-1}} - \dfrac{2}{z-1} \right).
\end{equation}
For $n = 2.08$, we get $\alpha = 0.33$ and $\theta = 0.5$. Plugging in $\theta = 0.5$ into Eq. (\ref{eq:xi nonrandom}) and solving Eq. (\ref{eq:fixedptEq}) yields the transmission coefficient as a function of $c$. Fig. \ref{fig:nonrand} shows the agreement between the physically rigorous empirical distribution and the analytically predicted distribution. Note in particular, the agreement for $c=2$ where the distribution is far from the characteristically bimodal DMPK distribution.

\begin{figure}
\centering
\subfloat[C=2.]{
\includegraphics[trim = 22 0 30 20, clip=true,width=0.45\textwidth]{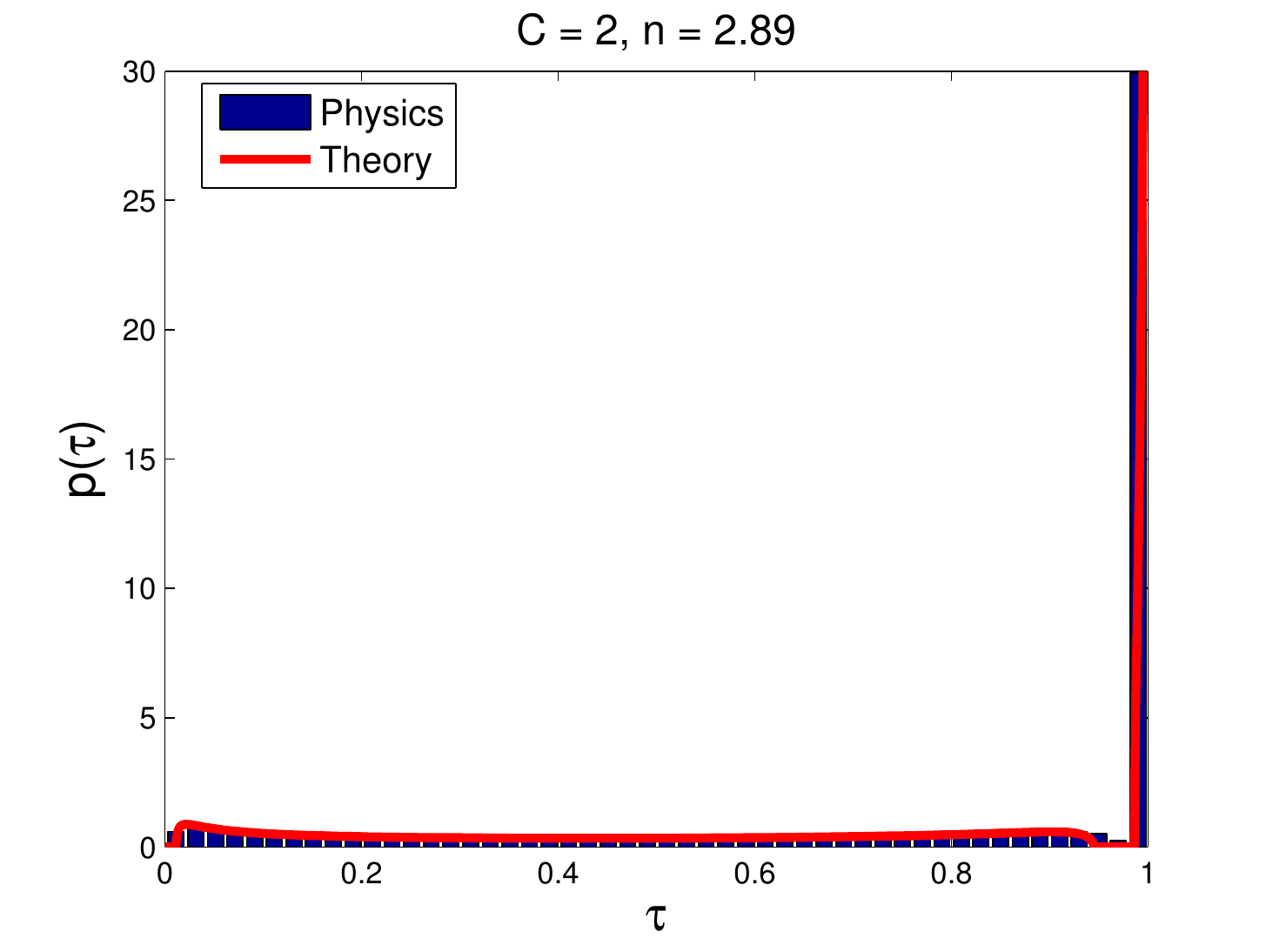}
}
\subfloat[C=33.]{
\includegraphics[trim = 22 0 30 20, clip=true,width=0.45\textwidth]{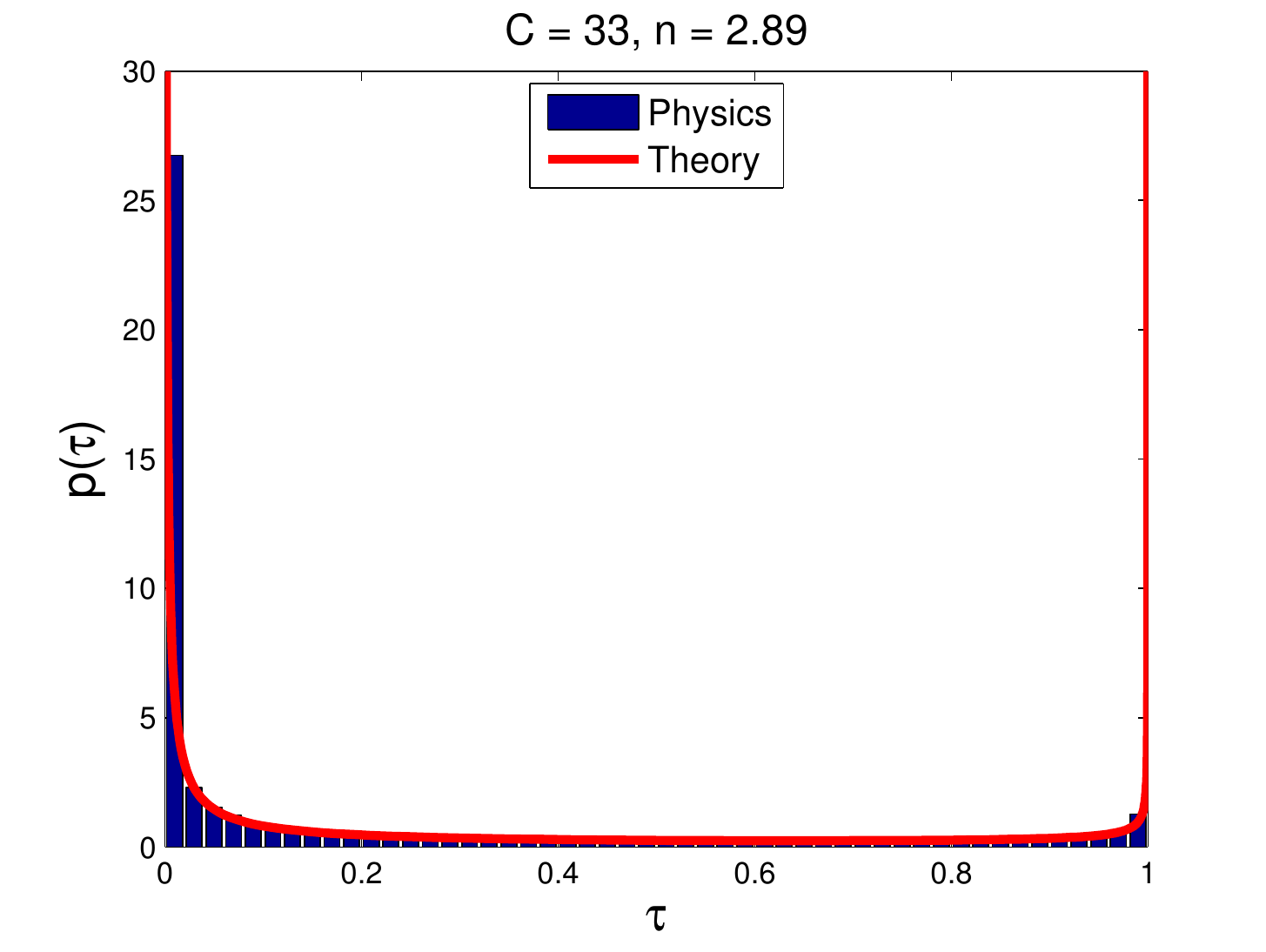}
}\\
\subfloat[C=2: Zoom-in. The predicted distribution is supported on two intervals.]{
\includegraphics[trim = 40 0 40 200, clip=true,width=0.90\textwidth]{PhysicsTheoryRandK1C2.pdf}
}\\
\subfloat[C=33: Zoom-in.]{
\includegraphics[trim = 40 0 20 200, clip=true,width=0.90\textwidth]{PhysicsTheoryRandK1C33.pdf}
}
\caption{The transmission coefficient distribution for the setting where $f_{\theta}(t) = 0.8 \delta(t-0.9) + 0.2 \delta(t-0.1)$. The red line is the theoretical prediction - the histograms are from the physically rigorous simulation averaged over $100$ trials. Note the agreement with theory in the $C=2$ where the distribution is far from the DMPK distribution.}
\label{fig:atomic}
\end{figure}

We now consider the setting where with probability $p_1$ a cylinder has refractive index $n_1$ and with probability $p_2$ it has a refractive index $n_2$. Plugging in $f_{\theta}(t) = p_1\delta(t-\theta_1) + p_2\delta(t-\theta_2)$ into Eq. (\ref{eq:xi general}) yields the expression
\begin{equation}\label{eq:xi two atom}
\widetilde{\xi}(z) = z\left( \dfrac{p_{1}}{z-\theta_{1}} + \dfrac{p_{1}}{z-\theta_{1}^{-1}} + \dfrac{p_{2}}{z-\theta_{2}} + \dfrac{p_{2}}{z-\theta_{2}^{-1}} - \dfrac{2}{z-1} \right).
\end{equation}
For $n_1 = 1.28$ and $n_2 = 2.89$ we get $\alpha_1 = 0.05$, $\theta_1 = 0.9$, and $\alpha_2 = 0.82$, $\theta_2 = 0.1$. Plugging in these values into Eq. (\ref{eq:xi two atom}) with $p_1 = 0.8$ and $p_2=0.2$ and solving Eq. (\ref{eq:fixedptEq}) yields the transmission coefficient as a function of $c$. Fig. \ref{fig:atomic} shows the agreement between the numerically obtained empirical distribution and the analytically predicted distribution. Note in particular, the agreement for $c=2$ where the predicted distribution is supported on two intervals and  agrees with the empirical results.

\begin{figure}
\centering
\subfloat[C=2.]{
\includegraphics[trim = 22 0 30 20, clip=true,width=0.45\textwidth]{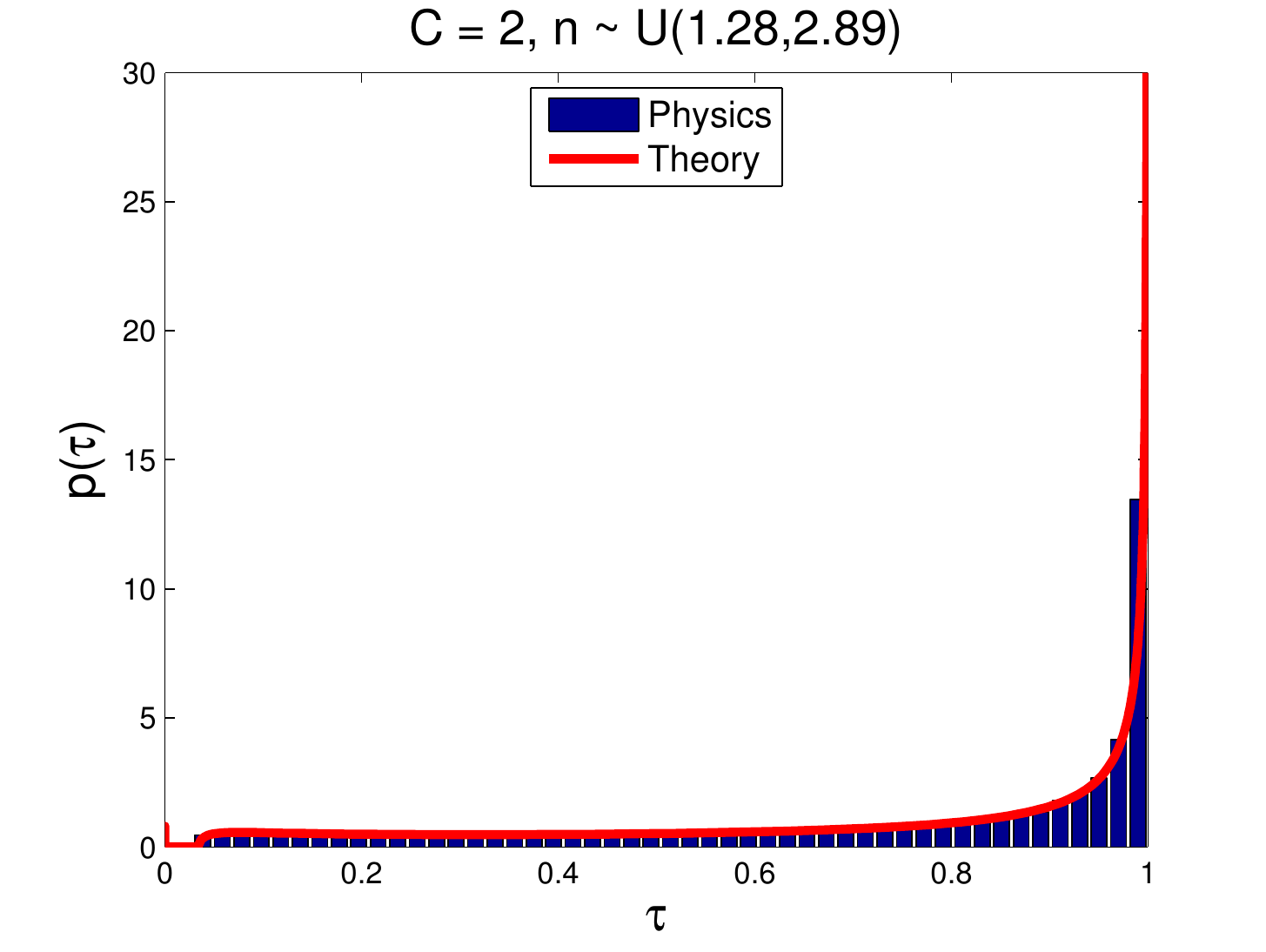}
}
\subfloat[C=25.]{
\includegraphics[trim = 22 0 30 20, clip=true,width=0.45\textwidth]{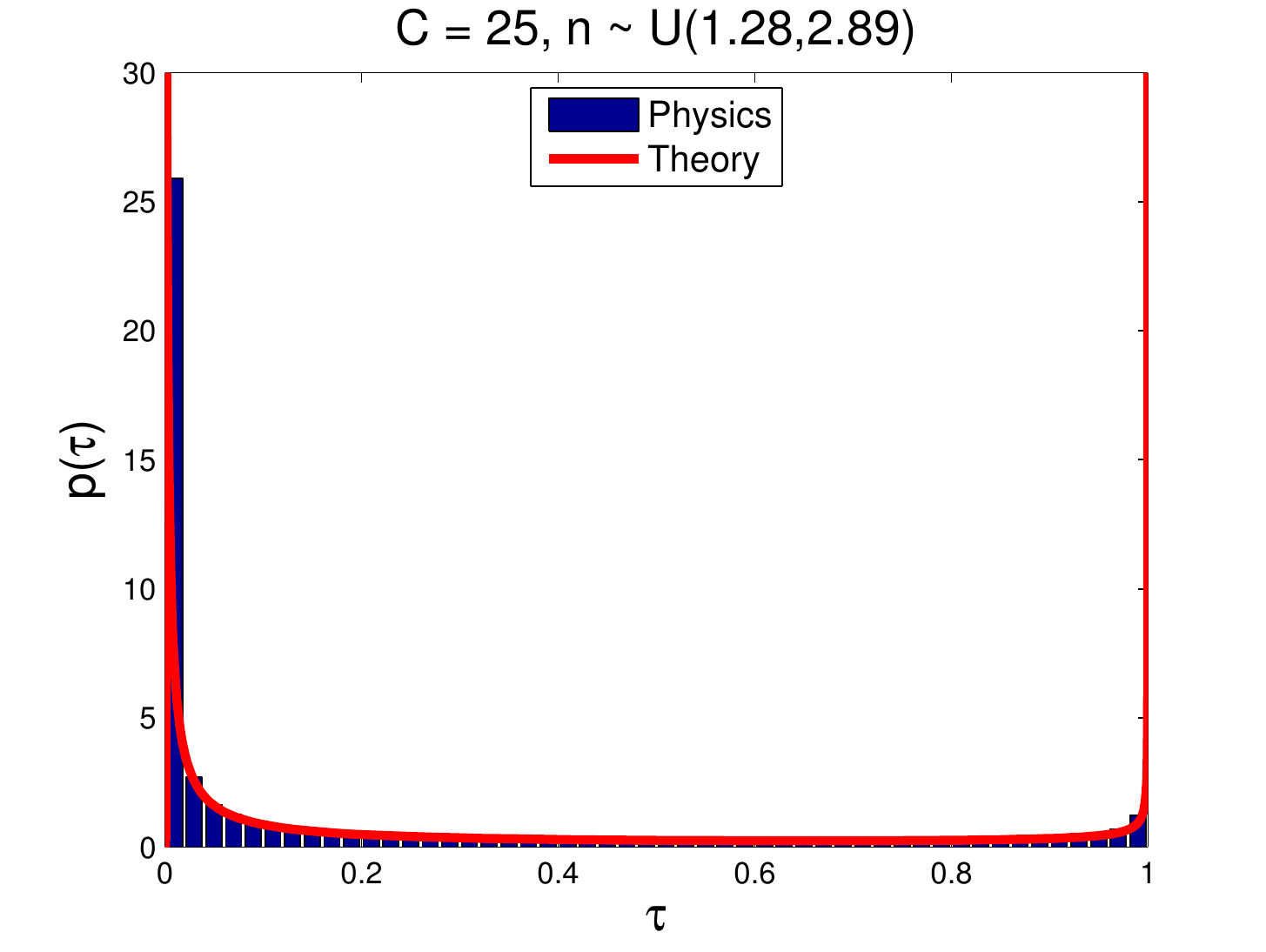}
}\\
\subfloat[C=2: Zoom-in.]{
\includegraphics[trim = 40 0 30 200, clip=true,width=0.90\textwidth]{PhysicsTheoryUniformK1C2.pdf}
}\\
\subfloat[C=25: Zoom-in.]{
\includegraphics[trim = 40 0 30 200, clip=true,width=0.90\textwidth]{PhysicsTheoryUniformK1C25.pdf}
}
\caption{The transmission coefficient distribution for the setting where $f_{\theta}(t) = \dfrac{1}{\theta_2-\theta_1} \mathbb{I}_{\theta_1 \leq t \leq \theta_2}$. The red line is the theoretical prediction - the histograms are from the physically rigorous simulation averaged over $100$ trials. Note the agreement with theory in the $C=2$ where the distribution is far from the DMPK distribution.}
\label{fig:uniform}
\end{figure}

Finally, we consider the setting corresponding to $f_{\theta}(t) = \dfrac{1}{\theta_2-\theta_1} \mathbb{I}_{\theta_1 \leq t \leq \theta_2}$. We generated the scattering system by mapping each random realization of $\theta$ to a random realization of the refractive index. Plugging this choice into Eq. (\ref{eq:xi general}) yields the expression
\begin{equation}\label{eq:xi uniform}
\widetilde{\xi}(z) = \dfrac{z}{\theta_{2}-\theta_{1}} \left( \log\left(\dfrac{z-\theta_{1}}{z-\theta_{2}}\right) + \dfrac{\theta_{2}-\theta_{1}}{z} + \dfrac{1}{z^2}\log\left( \dfrac{\theta_{2} z - 1}{\theta_{1} z -1} \right) - \dfrac{2}{z-1} \right).
\end{equation}
The choice of $\theta_1 = 0.1$ and $\theta_2 = 0.9$ corresponds to a refractive index of $n_1 = 2.89$ (with $\alpha_1 = 0.82$) and a refractive index $n_2 = 1.28$ (with $\alpha_2 = 0.9$). Plugging these values of $\theta_1$ and $\theta_2$ into Eq. (\ref{eq:xi uniform}) and solving Eq. (\ref{eq:fixedptEq}) yields the transmission coefficient as a function of $c$. Fig. \ref{fig:uniform} shows the agreement between the physically rigorous empirical distribution and the analytically predicted distribution.  Note in particular, the agreement for $c=2$ where the distribution is far from the characteristically bimodal DMPK distribution.

Appendix \ref{movies} contains some movies that shows the evolution of the transmission coefficient distribution with $c$ for each of the three scenarios discussed. As expected, for large enough $c$ the distribution eventually becomes characteristically bimodal as predicted by the DMPK theory. The behavior for small values of $c$ is accurately predicted by our theory.

For the three settings described above, we analytically compute $\mathbb{E}[\tau]$ from the associated $f_\theta$ via Eq. (\ref{eq:first moment closed form}). The computation involves the normalizing factor $B_2$, which for the three settings is given by
\begin{subequations}
\begin{align}\label{eq:B2 analytical}
B_{2}^{\rm nonrandom} &= \left( \dfrac{1-\theta}{1+\theta} \right)^2,\\
B_{2}^{\rm atomic} &= p_1 \left( \dfrac{1-\theta_1}{1+\theta_1} \right)^2 + p_2 \left( \dfrac{1-\theta_2}{1+\theta_2} \right)^2,\\
B_{2}^{\rm uniform} &= 1 - \dfrac{4}{\theta_2-\theta1}\log\left( \dfrac{\theta_2+1}{\theta_1+1} \right) + \dfrac{4}{(\theta_{1}+1)(\theta_{2}+1)}.
\end{align}
\end{subequations}
The closed-from expression for $\mathbb{E}[\tau^2]$ is lengthy and therefore omitted here. It can be obtained using the calculations in Appendix \ref{sec:second moment}.
Fig. \ref{fig:moment sims} compares the empirical moments with the predicted moments and shows the good agreement for a range of values of $c$.

\begin{figure}
\centering
\includegraphics[trim = 0 0 0 0, clip=true,width=0.95\textwidth]{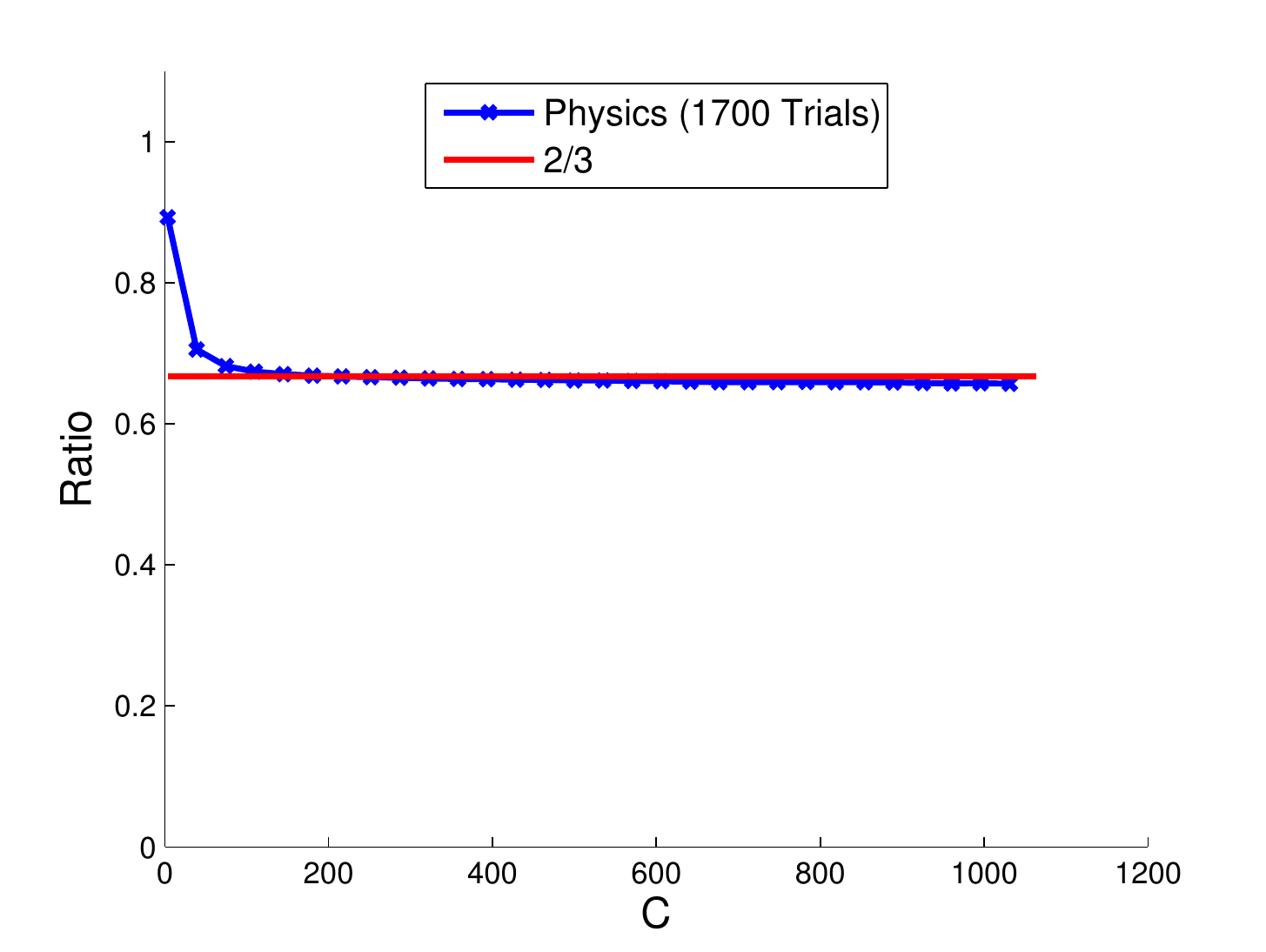}
\caption{The ratio of $\mathbb{E}[\tau^2]/\mathbb{E}[\tau]$ as a function of $c$. The $2/3$ line corresponds to the prediction in Eq. (\ref{eq:limiting ratio}) for the large $c$ limit of this ratio.}
\label{fig:ratio}
\end{figure}

Finally, we numerically validate the analytical prediction in Eq. (\ref{eq:limiting ratio}). To that end, we generated a random scattering system with $D=197\lambda, r=0.11\lambda, \widetilde{L} = 3.4 \times 10^5 \lambda, N_{c} = 430,000, n_{d} = 1.3$, and $M = 395$. The locations of the scatterers were selected randomly and produced a system with $\overline{l}=6.69\lambda$, where $\overline{l}$ is the average distance to the nearest scatterer. Let $L$ denote the thickness of the scattering system we are interested in analyzing. We vary $L$ from $\lambda$ to $\widetilde{L}$ and for each value of $L$ we compute the scattering matrices associated with only the scatterers contained in the $(0,L)$ portion of the $(0,\widetilde{L})$ system we have generated. This construction ensures that the average density per ``layer'' of the medium is about the same. We computed the first and second moment of the empirical transmission coefficient distribution by averaging over $1700$ random realizations of the scattering system and computed ratio as a function of $c = M/\widetilde{L}$. Fig. \ref{fig:ratio} shows that the empirical result validate our theoretical prediction.

\section*{Acknowledgements}
This work supported by  a DARPA Young Faculty Award  D14AP00086, AFOSR Young Investigator Award FA9550-12-1-0266, ONR Young Investigator Award N00014-11-1-0660, US Army Research Office (ARO) under grant W911NF-11-1-0391 and  NSF grant CCF-1116115.

\begin{figure}
\subfloat[The first moment.]{
\includegraphics[trim = 0 0 0 0, clip=true,width=0.45\textwidth]{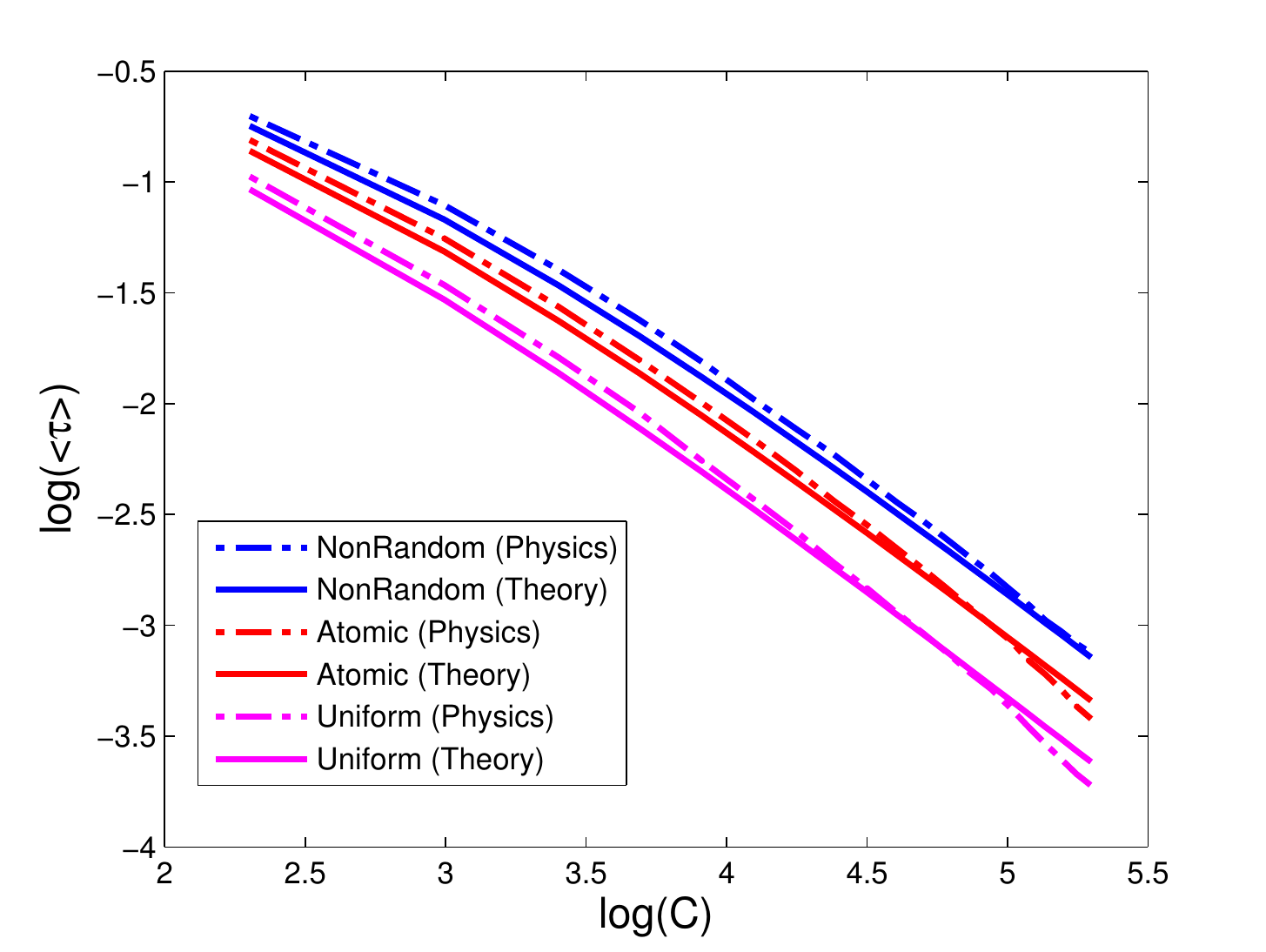}
}
\subfloat[The second moment.]{
\includegraphics[trim = 0 0 0 0, clip=true,width=0.45\textwidth]{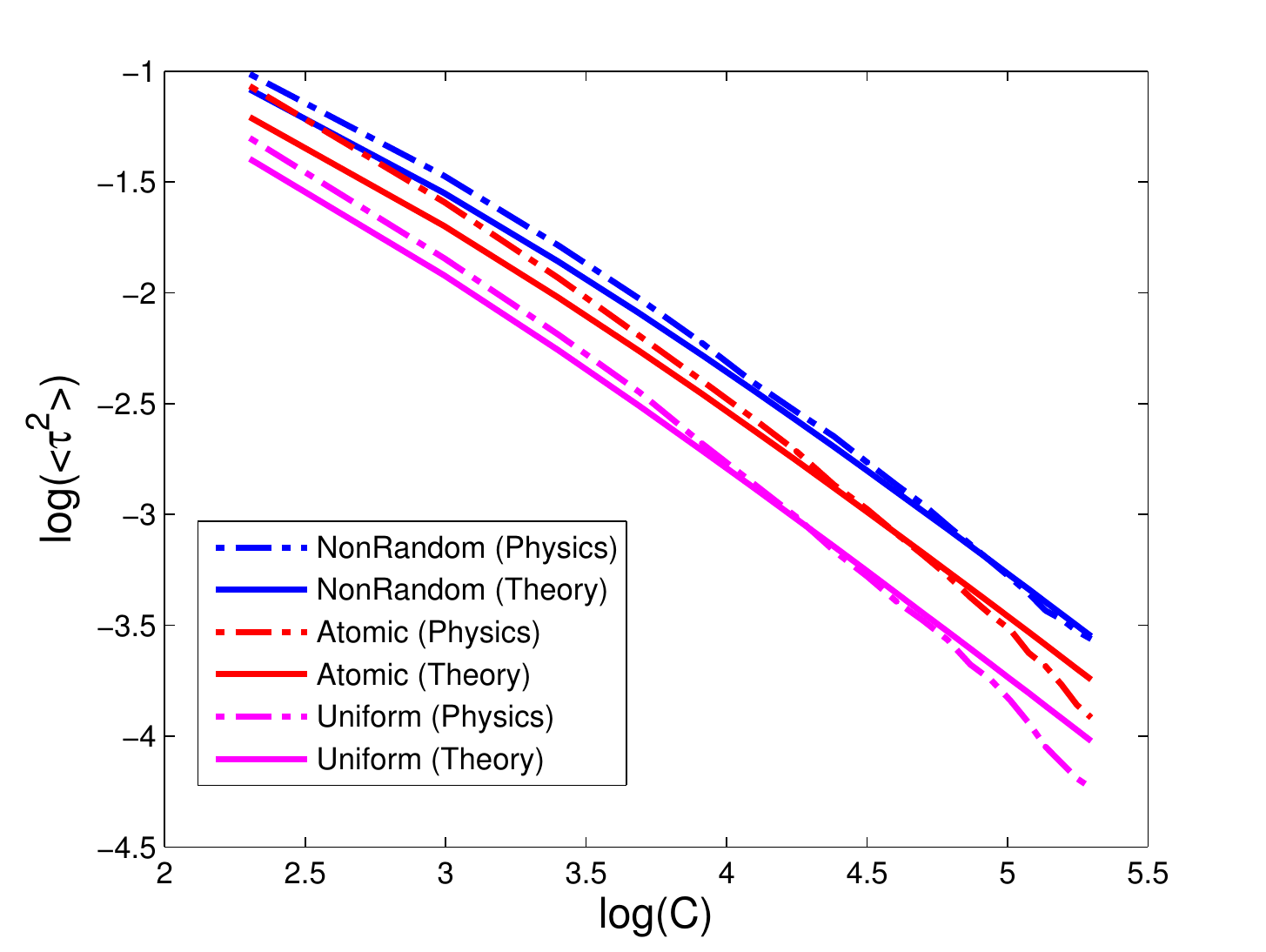}
}
\caption{The first moment versus $c$ for the settings corresponding to Fig. \ref{fig:nonrand}, Fig. \ref{fig:atomic} and Fig. \ref{fig:uniform} respectively.  The results of the physical simulations were averaged over $100$ trials.
}\label{fig:moment sims}
\end{figure}

\appendix
\section*{Appendices}
\section{Derivation of Eq. (\ref{eq:tth eigs})}\label{sec:appendix transfer}
Here, we uncover the relationship between the singular value squared of $S_{21}$, $\tau$, and the singular value squared of $T$, $\lambda$. Our derivation follows the approach in \cite[Section 1.C.1]{beenakker1997random}.
\subsection{Decomposition of $T^{H}\cdot T$}
Recall that the eigenvalues of $T^{H} \cdot T$ equal the square singular values of $T$. Using Eq. (\ref{eq:Tmatrix block}) and the fact that $S^{H} \cdot S = I$, we can express $T^{H}\cdot T$ as
\begin{align}
  \nonumber T^{H}\cdot T &=  \left[  \begin{array}{cc} S_{21}^{H}-S_{11}^{H}\cdot S_{12}^{-H}\cdot S_{22}^{H} & -S_{11}^{H}\cdot S_{12}^{-H}\\ S_{12}^{-H}\cdot S_{22}^{H} & S_{12}^{-H}\end{array} \right] \cdot  \left[  \begin{array}{cc} S_{21}-S_{22}\cdot S_{12}^{-1}\cdot S_{11} & S_{22}\cdot S_{12}^{-1}\\ -S_{12}^{-1}\cdot S_{11} & S_{12}^{-1}\end{array} \right] \\
  \label{app.THT} &= \left[ \begin{array}{cc} I + 2S_{11}^{H}\cdot S_{12}^{-H}\cdot S_{12}^{-1} \cdot S_{11}  & -2S_{11}^{H} \cdot S_{12}^{-H}\cdot S_{12}^{-1} \\ -2S_{12}^{-H} \cdot S_{12}^{-1}\cdot S_{11} & 2S_{12}^{-H} \cdot S_{12}^{-1} - I \end{array} \right].
\end{align}
In order to factorize the matrix on the right-hand side of Eq. (\ref{app.THT}) further, we first factorize the submatrices of the scattering matrix as
\begin{subequations}
\label{eq:factorization}
\begin{align}
  S_{21} &= U \cdot \Sigma \cdot V^{H}\\
  S_{11} &= F \cdot V^{*} \cdot \sqrt{I-\Sigma^{2}} \cdot V^{H}\\
  S_{12} &= F \cdot S_{21}^{T} \cdot F = F \cdot V^{*} \cdot \Sigma \cdot (F\cdot U^{*})^{H}\\
  S_{22} &= U \cdot \Ft \cdot \sqrt{I-\Sigma^{2}} \cdot (F\cdot U^{*})^{H},
\end{align}
\end{subequations}
where $\Ft = \Diag(\{ e^{j\phi_{n}} \}_{n})$ and $\phi_{n} \in [0,2\pi]$, and $\Ft$ represents the phase ambiguity between the singular spaces. Note that the factorizations in Eq. (\ref{eq:factorization}) satisfies power conservation, reciprocity and time-reversal symmetry. Substituting these into Eq. (\ref{app.THT}) yields the factorization
\begin{align}\label{eq:ThT factor}
   T^{H} \cdot T &= \left[ \begin{array}{cc} V \cdot (2\Sigma^{-2}-I) \cdot V^{H} & -2V \cdot \sqrt{I-\Sigma^{2}}\cdot \Sigma^{-2} \cdot (F\cdot V^{*})^{H} \\ -2F\cdot V^{*} \cdot \sqrt{I-\Sigma^{2}}\cdot \Sigma^{-2} \cdot V^{H} & F \cdot V^{*} \cdot (2\Sigma^{-2}-I) \cdot (F \cdot V^{*})^{H} \end{array} \right]\\
  \label{app.THTDecomp} &= \left[ \begin{array}{cc} V & 0 \\ 0 & F \cdot V^{*} \end{array} \right] \cdot \underbrace{\left[ \begin{array}{cc} 2\Sigma^{-2}-I & -2\sqrt{I-\Sigma^{2}} \cdot \Sigma^{-2} \\ -2\sqrt{I-\Sigma^{2}} \cdot \Sigma^{-2} & 2\Sigma^{-2} - I \end{array} \right]}_{=:\widetilde{\Sigma}} \cdot \left[ \begin{array}{cc} V & 0 \\ 0 & F \cdot V^{*} \end{array} \right]^{H}.
\end{align}
Note that this factorization reveals that the eigenvalues of $T^H\cdot T$ are exactly equal to the eigenvalues of $\widetilde{\Sigma}$.

\subsection{Eigenvalues of a special block matrix}
Note that the matrix $\widetilde{\Sigma}$ on the right-hand side of Eq. (\ref{eq:ThT factor}) is of the form
$$\left[ \begin{array}{cc} D_{1} & D_{2} \\ D_{3} & D_{4} \end{array} \right],$$
where $D_{1} = \Diag(\{d_{1,i}\}_{i=1}^{M})$, $D_{2} = \Diag(\{d_{2,i}\}_{i=1}^{M})$, $D_{3} = \Diag(\{d_{3,i}\}_{i=1}^{M})$ and $D_{4} = \Diag(\{d_{4,i}\}_{i=1}^{M})$. The eigenvalues $z$ of this block matrix are the solutions of the characteristic equation\begin{align*}
  \det\left(\left[ \begin{array}{cc} D_{1} - z I & D_{2} \\ D_{3} & D_{4} - z I \end{array}  \right]\right) = 0.
\end{align*}
Since the eigenvalues will not be the same as $d_{1,i}$, $D_{1}-z I$ will be invertible; hence the characteristic equation can be rewritten as
$$  \det(D_{1} - z I) \cdot \det(D_{4}-z I - D_{3} \cdot (D_{1} - z I)^{-1} \cdot D_{2}) = 0.$$
Equivalently,
$$\prod_{i=1}^{M}(d_{1,i}-z) \cdot \prod_{i=1}^{M}\left( d_{4,i}-z - \dfrac{d_{2,i}d_{3,i}}{d_{1,i}-z} \right) = \prod_{i=1}^{M}\left\{ z^{2} - (d_{1,i}+d_{4,i}) z + d_{1,i}d_{4,i} - d_{2,i}d_{3,i} \right\} =0.$$
Consequently, the $2M$ eigenvalues $z$ are given by
\begin{align} \label{app.eigD} z = \dfrac{d_{1,i}+d_{4,i} \pm \sqrt{(d_{1,i}+d_{4,i})^{2} - 4(d_{1,i}d_{4,i}-d_{2,i}d_{3,i})}}{2}\quad \mbox{for $i=1,\ldots,M$}.\end{align}

\subsubsection{Relationship between the singular values of $T$ and $S_{21}$}
Recall that $\tau_i = \sigma_i^{2}$ is the eigenvalue of $\Sigma^2$ and $\lambda_i$ is an eigenvalue of $T^{H}\cdot T$. We can apply Eq. (\ref{app.eigD}) to the matrix $\widetilde{\Sigma}$ in Eq. (\ref{app.THTDecomp}) to obtain the eigenvalues of $T^{H}\cdot T$, which are given by
\begin{align}
  \lambda_{i} &= \dfrac{(2/\tau_{i}-1)+(2/\tau_{i}-1) \pm \sqrt{(4/\tau_{i}-2)^{2}-4((2/\tau_{i}-1)^{2} -4(1-\tau_{i})/\tau_{i}^{2} )} }{2} \\
  &= 2/\tau_{i} -1 \pm 2\sqrt{1/\tau_{i}^{2} - 1/\tau_{i}}
\end{align}
Note that $1/\lambda_{i} = 2/\tau_{i} -1 \mp 2\sqrt{1/\tau_{i}^{2} - 1/\tau_{i}}$. This tells us that the singular values of the transfer matrix come in reciprocal pairs. Consequently, if we specify the singular values above one then the singular values below one are given by their reciprocal.

\section{Derivation of Eq. (\ref{eq:first moment closed form}) for $\mathbb{E}[\tau]$}\label{sec:first moment}
From Eq. (\ref{eq:mean2}) the first moment is given as \begin{equation*}
  \mathbb{E}[\tau] = -4\, \xi_{h}'(-1).
\end{equation*}
In order to evaluate this further, we are going to use what we have driven in Eq. (\ref{eq:psi final}),
\begin{align*}
  \psi_{h}(z; c) & = \exp\left( -c \cdot \dfrac{\widetilde{\xi}(1+\tfrac{1}{z}) }{z(z+1)}\right).
\end{align*}
Using the relationship between $\psi_{h}(z)$ and $\xi_{h}(z)$ in Eq. (\ref{eq:S transform}), we get\begin{equation}\label{eq:PsiXiRelation}
  \xi_{h}^{-1}(z) = \dfrac{1+z}{z}\dfrac{1}{\psi_{h}(z)} = \dfrac{1+z}{z} \exp\left( c \cdot \dfrac{\widetilde{\xi}(1+\tfrac{1}{z}) }{z(z+1)}\right).
\end{equation}
Therefore, \begin{equation}\label{eq:mean3}
  \mathbb{E}[\tau] = -4\, \xi_{h}'(-1) = -4\, \dfrac{1}{\partial_{z}\xi_{h}^{-1}(z)|_{z=z^{*}}}
\end{equation}
where $z^{*}$ is a value that satisfies $\xi_h(-1)=z^{*}$ or $\xi^{-1}_{h}(z^{*})=-1$. We can easily check that $z^{*} = -0.5$ by plugging it into Eq. (\ref{eq:PsiXiRelation}). \begin{align}
  \xi^{-1}_{h}(-0.5) = \dfrac{1-0.5}{-0.5}\exp\left( c \cdot \dfrac{\widetilde{\xi}(1+\tfrac{1}{-0.5}) }{-0.5(-0.5+1)}\right) = -1\exp\left( -4c \cdot \widetilde{\xi}(-1) \right).
\end{align}
From Eq. (\ref{eq:xi general}), we can evaluate $\widetilde{\xi}(-1)$ as follows \begin{align}
    \widetilde{\xi}(-1) &= \displaystyle - \int \left[ \dfrac{0.5}{-1-t} +  \dfrac{0.5}{-1-1/t} -\dfrac{1}{-1-1}  \right] f_{\theta}(t) dt\\
    &= \displaystyle - \int \left[ \dfrac{-0.5}{t+1} +  \dfrac{-0.5t}{t+1} +0.5  \right] f_{\theta}(t) dt = 0
\end{align}
Therefore we confirm\begin{align}
  \xi^{-1}_{h}(-0.5) = -1\exp\left( -4c \cdot \widetilde{\xi}(-1) \right) = -1.
\end{align}

To complete Eq. (\ref{eq:mean3}), we need to evaluate $\partial_{z}\xi_{h}^{-1}(z)|_{z=-0.5}$. Straightforward evaluation leads to the following result,\begin{equation}
\label{eq:xiInv1}
\partial_{z}\xi_{h}^{-1}(z)|_{z=-0.5} = -4\left(1+c\, \displaystyle\int \left( \dfrac{1- t}{1 +t} \right)^2 f_{\theta}(t) dt\right)
\end{equation}
and we get
$$ \mathbb{E}[\tau] = -4\, \xi_{h}'(-1) = -4\, \dfrac{1}{\partial_{z}\xi_{h}^{-1}(z)|_{z=-0.5}} = \dfrac{1}{1+c\, \displaystyle\int \left( \dfrac{1- t}{1 +t} \right)^2 f_{\theta}(t) dt }.$$

\section{Closed-form expression for $\mathbb{E}[\tau^2]$}\label{sec:second moment}

For notational convenience we define

\begin{subequations}
\begin{align}
  B_{n} &:= \int \left( \dfrac{1- t}{1 +t} \right)^n f_{\theta}(t) dt \label{eq:Bn}\\
  \xi_{h}^{-1'} &:= \left. \partial_{z} \xi_{h}^{-1}(z) \right|_{z=-0.5}\\
  \xi_{h}^{-1''} &:= \left. \partial^{2}_{z} \xi_{h}^{-1}(z) \right|_{z=-0.5}\\
  \xi_{h}^{-1'''} &:= \left. \partial^{3}_{z} \xi_{h}^{-1}(z) \right|_{z=-0.5}.
\end{align}
\end{subequations}

From Eq. (\ref{eq:second moment}) we have
\begin{equation}\label{eq:Etau sq calc}
 \mathbb{E}[\tau^{2}] = 16 \left[ \dfrac{1}{6}  \xi_{h}'''(-1) - \dfrac{1}{2} \xi_{h}''(-1)  \right].
 \end{equation}
As we did for the first moment, we are going to express this in terms of the inverse function of $\xi(z)$. We are going to use the following elementary results from calculus
\begin{subequations}
\begin{align}\label{eq:elem calculus}
  \xi_{h}''(-1) &= -\dfrac{\xi_{h}^{-1''}}{\left( \xi_{h}^{-1'} \right)^3}\\
  \xi_{h}'''(-1) &= \dfrac{3\left( \xi_{h}^{-1''} \right)^2 - \xi_{h}^{-1'} \xi_{h}^{-1'''}}{\left(\xi_{h}^{-1'}\right)^5}.
\end{align}
\end{subequations}

Substituting Eq. (\ref{eq:elem calculus}) into Eq. (\ref{eq:Etau sq calc}) yields the expression
\begin{align}
 \nonumber \mathbb{E}[\tau^{2}] &= 16 \left[ \dfrac{1}{6}  \xi_{h}'''(-1) - \dfrac{1}{2} \xi_{h}''(-1)  \right]\\
 \label{eq:second moment2}
 &= \dfrac{8}{3}\dfrac{3(\xi_{h}^{-1''})^2 + 3\xi_{h}^{-1''} - \xi_{h}^{-1'} \xi_{h}^{-1'''}}{ (\xi_{h}^{-1'})^5 }
 \end{align}
Note that $\xi_{h}^{-1'}$,  $\xi_{h}^{-1''}$ and $\xi_{h}^{-1'''}$ can be expressed in terms of $B_n$, defined in Eq. (\ref{eq:Bn}) as
\begin{subequations}\label{eq:Invc}
\begin{align}
  \label{eq:xiInv1new} \xi_{h}^{-1'} &= -2^2 (1 + cB_2 )\\
  \label{eq:xiInv2} \xi_{h}^{-1''} &= -2^4 (1 + 2cB_2 + c^2B_4)\\
  \label{eq:xiInv3}\xi_{h}^{-1'''} &= -2^5 \left(3 + 6c B_2 + 6(c + c^2)B_4 + 2c^3B_6\right).
\end{align}
\end{subequations}
The full (messy) expression for the second moment can be obtained by plugging Eqs. (\ref{eq:xiInv1new}), (\ref{eq:xiInv2}) and (\ref{eq:xiInv3}) into Eq. (\ref{eq:second moment2}).

\section{Movies showing evolution of the transmission coefficient distribution with $c$} \label{movies}
\centerline{\animategraphics[controls,buttonsize=0.3cm,width=12.5cm]
    {10}{"movies/EvolutionNonRand"}{1}{259}}
\centerline{\animategraphics[controls,buttonsize=0.3cm,width=12.5cm]
    {10}{"movies/EvolutionRand"}{1}{259}}
\centerline{\animategraphics[controls,buttonsize=0.3cm,width=12.5cm]
    {10}{"movies/EvolutionUniform"}{1}{259}}

\end{document}